\documentclass[notoc]{JHEP3}
\usepackage{graphicx,psfrag}
\usepackage{color,cite}
\let\normalcolor\relax
\newcommand{\eqref}[1]{(\ref{#1})}
\newcommand{\text}[1]{{\rm #1}}

\newenvironment{changemargin}[2]{%
  \begin{list}{}{%
    \setlength{\topsep}{0pt}%
    \setlength{\leftmargin}{#1}%
    \setlength{\rightmargin}{#2}%
    \setlength{\listparindent}{\parindent}%
    \setlength{\itemindent}{\parindent}%
    \setlength{\parsep}{\parskip}%
  }%
  \item[]}{\end{list}}

\newcommand{\be}{\begin{equation}}
\newcommand{\ee}{\end{equation}}
\newcommand{\cmb}{\begin{changemargin}}
\newcommand{\cme}{\end{changemargin}}
\newcommand{\bea}{\begin{eqnarray}}
\newcommand{\eea}{\end{eqnarray}}

\newcommand{\highlightA}{black}
\newcommand{\highlight}{black}

\def\spa#1.#2{\langle#1\,#2\rangle}
\def\spb#1.#2{[#1\,#2]}
\def\sandmm#1.#2.#3{%
\left\langle\smash{#1}{\rphantom1}\right|{#2}%
\left|\smash{#3}{\rphantom1}\right]}
\def\spab#1.#2.#3{\sandmm#1.#2.#3}
\def\spba#1.#2.#3{\sandpp#1.#2.#3}
\def\spaa#1.#2.#3.#4{\sandmp#1.{#2#3}.#4}
\def\spbb#1.#2.#3.#4{\sandpm#1.{#2#3}.#4}
\def\spash#1.#2{\spa{\smash{#1}}.{\smash{#2}}}
\def\spbsh#1.#2{\spb{\smash{#1}}.{\smash{#2}}}

\def\ksl{\not{\hbox{\kern-2.3pt $k$}}}

\title{Towards the understanding of jet shapes and cross sections in heavy ion collisions using soft-collinear effective theory}

\author{Yang-Ting Chien and Ivan Vitev\\
Theoretical Division, T-2\\
Los Alamos National Laboratory\\
Los Alamos, NM 87545, USA}

\abstract{
We calculate the jet shape and the jet cross section in heavy ion collisions using soft-collinear effective theory (SCET) and its extension with Glauber gluon interactions in the medium (SCET$_{\rm G}$). We use the previously developed framework to systematically resum the jet shape at next-to-leading logarithmic accuracy, and we consistently include the medium modification by incorporating the leading order medium-induced
splitting functions. The calculation provides, for the first time, a quantitative understanding of the jet shape modification measurement in lead-lead collisions at $\sqrt{s_{\rm NN}}=2.76$ TeV at the LHC. The inclusive jet suppression is also calculated within the same framework beyond the traditional concept of parton energy loss, and the dependence on the centrality, the jet radius and the jet kinematics is examined. In the end we present predictions for the anticipated jet shape and cross section measurements in lead-lead collisions at $\sqrt{s_{\rm NN}}\approx5.1$ TeV at the LHC.
}

\begin{document}
%%%%%%%%%%%%%%%%%%%%%%%%%%%%%%%%%%%%%%%%%%%%%%%%%%%
\section{Introduction}
\label{sec:intro}
The Run II of the Large Hadron Collider (LHC) has started this year, and the nucleon-nucleon center of mass energy of the lead-lead collisions is planned to be boosted above 5 TeV. More energetic jets will be produced, which make the LHC an excellent arena for the study of jet physics in heavy ion collisions. On the other hand, one of the top priorities of the heavy ion program at the LHC is to study the properties of the hot, dense medium that is produced during the collisions and referred to as the quark-gluon plasma (QGP). Many interesting questions are under investigation, for example, how the medium is created and subsequently thermalizes, what the dynamics of the medium is and how the medium properties evolve with energy and density. To answer these questions, one of the key insights is to study how the presence of the medium affects various hard processes as probes of the QGP. The medium responses to the perturbation from the hard probes may also contain useful information.

With the abundance of jets and the high-resolution detectors at the LHC, one promising approach to extract the medium properties is to study the modification of jets as they propagate through the medium. This is the jet quenching \cite{Gyulassy:1990ye}
%\cite{Bjorken:1982tu,Gyulassy:1990ye}
phenomenum which has been observed at the Relativistic Heavy Ion Collider (RHIC) \cite{Adcox:2001jp,Adler:2002xw,Adcox:2004mh,Arsene:2004fa,Back:2004je,Adams:2005dq} and the LHC \cite{CMS:2012aa,Aamodt:2010jd,Abelev:2012hxa,Abelev:2013kqa,Aad:2012vca,Aad:2014bxa,Chatrchyan:2013kwa,CMS:prelim,
Chatrchyan:2012gw,Chatrchyan:2014ava,Aad:2014wha,Chatrchyan:2013exa,
Adam:2015ewa,Chatrchyan:2012gt,Chatrchyan:2011sx,Chatrchyan:2012nia,Aad:2010bu,Aad:2013sla,Adam:2015doa,Aad:2015bsa}. Traditionally, jet quenching refers to the suppression of production cross sections for various hadrons. The suppression was understood theoretically with several different approaches \cite{Gyulassy:1993hr,Wang:1994fx,Zakharov:1996fv,Zakharov:1997uu,Baier:1996kr,Baier:1998kq,
Gyulassy:2000er,Gyulassy:2000fs,Wiedemann:2000za,Wang:2001ifa,
Arnold:2001ba,Arnold:2001ms,Arnold:2002ja,Casalderrey-Solana:2014bpa} (for reviews with different perspectives, see \cite{Gyulassy:2003mc,Majumder:2010qh,Mehtar-Tani:2013pia,CasalderreySolana:2011us}). They share the common physical picture of the parton energy loss, but with emphasis on physics at different energy scales. It has been clear that hadron cross section suppression alone is not enough to identify and distinguish between different physical mechanisms in jet quenching \cite{Bass:2008rv,Renk:2011aa,Armesto:2011ht}, and more differential and correlated measurements are needed~\cite{Vitev:2008rz,Gavai:2015pka}.

The studies of jet substructure observables and production cross section can provide precision tests of the QCD dynamics. Historically, the studies of event shapes and cross sections in $e^+e^-$ collisions helped confirm the gauge theory structure of QCD \cite{Sterman:1977wj,Farhi:1977sg, Georgi:1977sf, PhysRevLett.41.1581, PhysRevLett.41.1585, PhysRevD.19.2018, Heister:2003aj, Abdallah:2003xz, Achard:2004sv, Abbiendi:2004qz}. The discovery of the asymptotic freedom of QCD at high energy allowed the understanding of jets using perturbative tools. Since then, accurate QCD calculations provide robust theory control of jet physics \cite{Ellis:1990ek,Ellis:1994dg,Ellis:1992en}. They have also allowed one of the most precise extraction of the strong coupling constant \cite{Becher:2008cf, Chien:2010kc, Davison:2008vx, Abbate:2010xh,Hoang:2015hka}. In heavy ion collisions, the intricate jet formation mechanism provides the sensitivity to the medium dynamics. It is expected that precision QCD calculations will be crucial in the understanding of jets, therefore allowing us to extract the medium properties reliably.

In general, jet substructure observables are more sensitive to the details of the final state, jet-medium interactions. They allow us to disentangle the initial state, cold nuclear matter (CNM) effects and therefore are cleaner probes of the medium. On the other hand, different jet substructure observables are sensitive to radiation at different energy scales. By measuring, for example, from jet cross sections, jet shapes and jet fragmentation functions to angularities, jet masses and particle multiplicities, the in-medium jet formation mechanism across a wide range of energy scales can be examined. Jet substructure observables and their medium modifications are also highly dependent on the partonic origin of jets. Quark-initiated and gluon-initiated jets typically have different substructures because of the different color charges \cite{Gallicchio:2011xq,Gallicchio:2012ez}. With different medium interactions, quark-jets and gluon-jets can even be considered independent probes.

It should be noted that, jets are conventionally defined by a jet algorithm used in the jet reconstruction. An angular scale $R$ is introduced, and jets are identified as exclusive objects in an event. While jet substructure observables probes directly the radiation inside jets, the jet cross section can give complimentary information about the radiation in the rest of the event. In heavy ion collisions, because of the huge underlying event background, a small radius is usually chosen in jet reconstruction. This leads to the significant suppression of the jet cross section because the radiation can easily be lost outside jets. Recent progress has been made on the precision calculations of jet cross sections for jets with small radii \cite{Dasgupta:2014yra,Chien:2015cka} and a better understanding of the radius dependence of the cross section.

Because of the sensitivity to physics at multiple energy scales, the theoretical calculations of jet substructure observables are challenging tasks, which has been realized for a long time. At high energy, perturbative calculations are reliable, and the contributions from non-perturbative regimes are power-suppressed. However, because of the emergence of hierarchical scales, the calculations of jet substructure observables always suffer from large logarithms of the ratio between these scales. The resummation of the perturbative series thus becomes necessary, and the medium modification of jet substructure observables will have to be consistently included.

We will study the jet shape and the jet cross section in heavy ion collisions. The jet shape \cite{Ellis:1992qq} is one of the classic jet substructure observables and it probes the transverse energy profile inside a jet. Given a jet of size $R$ reconstructed using a jet algorithm, the integral jet shape $\Psi_J(r)$ is defined as the fraction of the transverse energy $E_T$ of the jet within a subcone of size $r$ around the jet axis $\hat n$,
\be
    \Psi_J(r)=\frac{\sum_{i,~d_{i\hat n}<r} E^i_T}{\sum_{i,~d_{i\hat n}<R} E^i_T}\;.
\ee
Here $d_{i\hat n}=\sqrt{(\eta_i-\eta_{jet})^2+(\phi_i-\phi_{jet})^2}$ is the Euclidean distance between the $i$-th particle in the jet and the jet axis on the pseudorapidity-azimuthal angle $(\eta,\phi)$ plane. The average integral jet shape $\Psi(r)=\frac{1}{N_J}\sum_{J=1}^{N_J} \Psi_J(r)$ and its derivative, the differential jet shape $\rho(r)=\frac{d}{dr}\Psi(r)$, describe how the transverse energy inside jets is distributed in $r$. In heavy ion collisions, the modification of jet shapes is conventionally evaluated by the following ratio
\be
M_{\rho}(r) =  \frac{ \rho^{AA}(r)}{ \rho^{pp}(r) }\;.
\label{RAAshape}
\ee
For the modification of jet cross sections, we measure the nuclear modification factor $R_{AA}$ defined as follows,
\be
R_{AA} =
{ \frac{d\sigma_{AA}}{d\eta d p_T} } \Bigg/
{ \langle  N_{\rm bin}  \rangle
\frac{ d\sigma_{pp}}{d\eta d p_T}  } \;,
\label{RAAjet}
\ee
where $\langle  N_{\rm bin}  \rangle$ is the number of binary nucleon-nucleon collisions in an A+A reaction.

Historically, the jet shape was first resummed using the modified leading logarithmic approximation in \cite{Seymour:1997kj}, and later in \cite{Li:2011hy,Li:2012bw} using a different approach. In heavy ion collisions, the work \cite{Vitev:2008rz,Vitev:2009rd} built upon \cite{Seymour:1997kj} and studied the medium modification of jet shapes using the Gyulassy-Levai-Vitev formalism \cite{Gyulassy:2000er,Gyulassy:2000fs} in the soft gluon limit. Other approaches include \cite{Salgado:2003rv}, and Monte Carlo studies of jet shapes have also been performed \cite{Renk:2009hv,Ma:2013uqa,Ramos:2014mba}. On the other hand, calculations based on the framework of traditional perturbative parton energy loss~\cite{He:2011pd,Neufeld:2010fj,Neufeld:2012df,Dai:2012am} have been able to describe the suppression of light-flavor jets and its radius dependence, heavy flavor jets~\cite{Stavreva:2012aa,Huang:2013vaa,Huang:2015mva,Chatrchyan:2013exa}, as well as the enhanced asymmetry in di-jet and $\gamma/Z^0$+jet events observed in heavy ion collisions~\cite{Aad:2010bu,Chatrchyan:2011sx,Aad:2012vca,Chatrchyan:2012gt}. Other approaches incorporate energy redistribution in the parton shower through various transport models, typically implemented in Monte Carlo simulations and event generators~\cite{Lokhtin:2011qq,Young:2011qx,Qin:2012gp,Wang:2013cia,Schenke:2009gb,Armesto:2009fj,Zapp:2013vla,Zapp:2013zya}.

In \cite{Chien:2014nsa,Chien:2014zna}, we initiated the studies of precision jet substructure and cross section calculations in heavy ion collisions using soft-collinear effective theory (SCET) \cite{Bauer:2000ew, Bauer:2000yr, Bauer:2001ct, Bauer:2001yt, Bauer:2002nz,Beneke:2002ph} and its extension to include Glauber gluon interactions in the medium ($\rm SCET_G$) \cite{Idilbi:2008vm, Ovanesyan:2011xy}. The goal of such development is to incorporate recent advances in pQCD and SCET in accurate theoretical predictions of jet observables in heavy ion collisions. In this formalism, Glauber gluons are treated as background fields generated from the color charges in the QGP. They mediate the interactions between the collinear jet and the QCD medium. $\rm SCET_G$ allows us to calculate the medium-induced splitting functions \cite{Ovanesyan:2011kn,Fickinger:2013xwa} beyond the traditional soft-gluon emission limit, which have been applied to describe the suppression of charged-hadron and neutral-pion cross sections at RHIC and the LHC~\cite{Kang:2014xsa,Chien:2015vja}. In this paper, we will present the first perturbative calculation of the jet shape modification and jet cross section suppression beyond the traditional energy loss approach.

The rest of the paper is organized as follows. In Sec.~\ref{sec:fram} we present the theoretical framework for the calculation of the medium-modified jet shapes and cross sections. We demonstrate how the medium-induced splitting functions are incorporated naturally in the jet energy function and the jet energy loss calculations, with the implementation of cold nuclear matter effects. In Sec.~\ref{sec:results} we compare the theoretical calculations to the experimental measurements in Pb+Pb collisions at $\sqrt{s_{\rm NN}} = 2.76$~TeV at the LHC. We also show predictions for the upcoming $\sqrt{s_{\rm NN}} \approx 5.1$~TeV measurements. In Sec.~\ref{sec:conc} we summarize and give an outlook of future studies.

\section{Framework}
\label{sec:fram}

With the relevance of multiple energy scales in heavy ion jet physics, we aim to develop an effective field theory framework in which both the resummation of jet substructure observables, as well as the calculation of their medium modifications, can be performed consistently, and the precision can be systematically improvable. In this section we review the recently developed framework in SCET \cite{Bauer:2000ew, Bauer:2000yr, Bauer:2001ct, Bauer:2001yt, Bauer:2002nz} for the resummation of jet shapes in proton collisions \cite{Chien:2014nsa}. For the jet shape in heavy ion collisions, this framework is readily generalized \cite{Chien:2014zna} to include medium modifications via Glauber gluon interactions \cite{Idilbi:2008vm, Ovanesyan:2011xy}. The medium-induced splitting functions \cite{Ovanesyan:2011kn,Fickinger:2013xwa} were calculated in $\rm SCET_G$ in a medium model consisting of thermal quasi-particles undergoing longitudinal Bjorken-expansions \footnote{\textcolor{\highlight}{
The medium-induced splitting functions depend on the density and the geometry of the medium. At the LHC the gluon density dominates and it is related to the soft, charged hadron rapidity density which is proportional to the number of participants in the optical Glauber model. The proportionality coefficient is extracted from the experimental measurements in lead-lead collisions up to $\sqrt{s_{\rm NN}} = 2.76$~TeV and extrapolated to $5.1$~TeV. \textcolor{\highlightA}{In particular, the gluon rapidity densities $dN/dy$ we use in the calculation are as follows: at 2.76 TeV, we use 2640 (750) for central (mid-peripheral) collisions, while at 5.1 TeV we use 2750 (800).} The medium-induced splitting functions are averaged within each centrality class over the spacetime position of the hard scattering process in the Glauber model, allowing all possible path lengths in the average. The event-by-event fluctuation of jet modifications and the correlation with the medium geometry will be studied in the future.
}}.
They aim at describing how the parton splitting in the full kinematic regime changes in a medium with the characteristic scale at the temperature $T$. The effect from a more subtle medium spacetime evolution is expected to be suppressed by the jet energy. \footnote{While many bulk properties of the medium require a complete hydrodynamic description, in this work we focus on jet physics above medium scales and evaluate the medium-induced splitting functions using the classic Bjorken-expansion model of the medium. Medium effects at lower energies on jets will be investigated in the future.}

SCET is an effective field theory of QCD suitable for describing processes producing jets with energetic collinear particles. In such processes, a dynamical scale $m_J$ emerges where $m_J$ is the jet mass and is much lower than the jet energy $Q$. SCET expands the full QCD contributions in the calculation with a systematic power counting. The power counting parameter $\lambda\sim m_J/Q$ is typically small and the leading-power contribution calculated in SCET is a good approximation. The power-suppressed corrections can also be included order by order systematically. In the calculation of certain types of jet substructure observables, an even lower scale $m_J^2/Q$ emerges which is the scale of the ubiquitous soft radiation in an event. SCET separates physics at these hierarchical scales and makes the factorization of the physical cross section manifest. This allows the calculation to be factorized into pieces, each can be performed at its natural scale more easily. The contribution in each sector at any other scales can then be obtained by the renormalization group evolution.

In SCET we have the following hierarchy of energy scales,
\be
    Q \gg Q\lambda \gg Q\lambda^2  \gtrsim   \Lambda_{\rm QCD}.
\ee
In the presence of the medium at the characteristic scale $T$, the soft scale $Q\lambda^2$ can be comparable to the medium temperature. For a jet along the direction $\hat n$,  in light-cone coordinates the momentum is expressed as $p=(\bar n\cdot p, n\cdot p, \vec p_\perp)$ where $n=(1,\hat n)$ and $\bar n=(1,-\hat n)$. The momentum scalings of the collinear and soft particles in the jet, as well as the Glauber modes mediating the medium interaction, are
\be
    p_c=Q(1,\lambda^2,\lambda)\;,~p_s=Q(\lambda,\lambda,\lambda)\;,~p_G=Q(\lambda^2,\lambda^2,\lambda)\;.
\ee
Note that the Glauber modes are off-shell degrees of freedom describing the momentum transfer transverse to the collinear direction. They are not final state particles in the jet and therefore do not directly contribute to the measurement of jet substructure observables.

In the splitting of a collinear parton with momentum $p=(\omega,0,0)$ into partons with momenta $k=(x\omega,k_\perp^2/x\omega,k_\perp)$ and $p-k$, the formation time of such splitting process is
\be
    t_{split}\sim \frac{x(1-x) \omega}{k_\perp^2} \sim \frac{1}{Q\lambda^2}.
\ee
This is to be compared with the typical mean free path of the parton multiple scattering in the medium,
\be
    t_{free}\sim \frac{1}{T},% \sim \left( \frac{1}{\lambda Q }  ,    \frac{1}{\lambda^2 Q } \right )   ,
\ee
and
%for collinear splitting
$t_{split}$ can be comparable to $t_{free}$ or even larger. This leads to the coherent multiple scattering of partons which underlies the Landau-Pomeranchuk-Migdal (LPM) effect. The induced radiation in the collinear direction is suppressed, resulting in the wide-angle bremsstrahlung~\cite{Vitev:2005yg} which is captured by the $\rm SCET_G$ medium-induced splitting functions.

In the context of the jet shape calculation, the integral jet shape has dominant contributions from the collinear sector in SCET,
\be
    \Psi(r)=\frac{E_r}{E_R}=\frac{E_r^c+E_r^s}{E_R^c+E_R^s}=\frac{E_r^c}{E_R^c}+{\cal O}(\lambda),
\ee
because $E^c = {\cal O}(Q)$ and $E^s = {\cal O}(Q\lambda)$. $E^c$ and $E^s$ are the contributions to the measured energies from collinear and soft radiation. The collinear nature of the jet energy leads to the simple factorized expression of the jet shape \cite{Chien:2014nsa},
\be
    \Psi_\omega(r)=\frac{J_{\omega,E_r}(\mu)}{J_{\omega,E_R}(\mu)}\;,
\ee
which is the ratio between two jet energy functions, and $\mu$ is the renormalization scale \textcolor{\highlight}{appearing in the calculation using dimensional regularization in $d$ dimensions and the $\overline{MS}$ renormalization scheme}. Here
\be
    J_{\omega,E_r}(\mu)=\sum_{X_c}\langle 0|\bar\chi_{\omega}(0)|X_c\rangle\langle X_c|\chi_{\omega}(0)|0\rangle \hat E_r(X_c)\;,
\ee
is the jet energy function which describes the average amount of jet energy inside the subcone of size $r$ radiated from the parton with the collinear momentum $\bar n\cdot p=\omega$. $\chi_\omega$ is the collinear jet field operator and $X_c$ is the collinear sector constrained within the angular scale $R$ used in the jet reconstruction. The measurement operator $\hat E_r$ returns the energy inside the subcone of size $r$. More detailed information about the above expression can be found in \cite{Chien:2014nsa}, nevertheless it simply calculates the jet energy weighted by the probability using SCET matrix elements.

\begin{figure}[!t!t]
\centering
\includegraphics[width=0.49\textwidth]{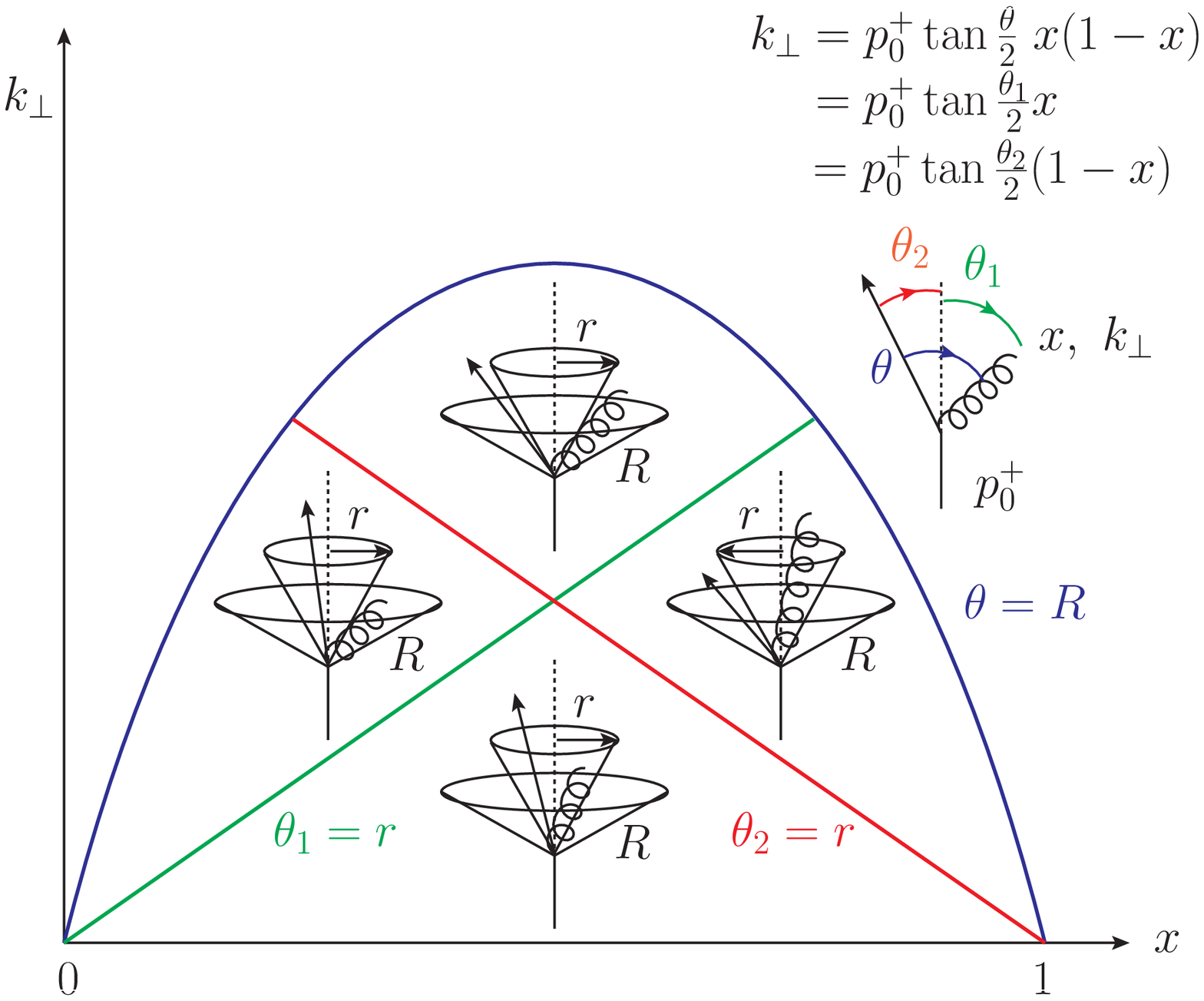}
\includegraphics[width=0.49\textwidth]{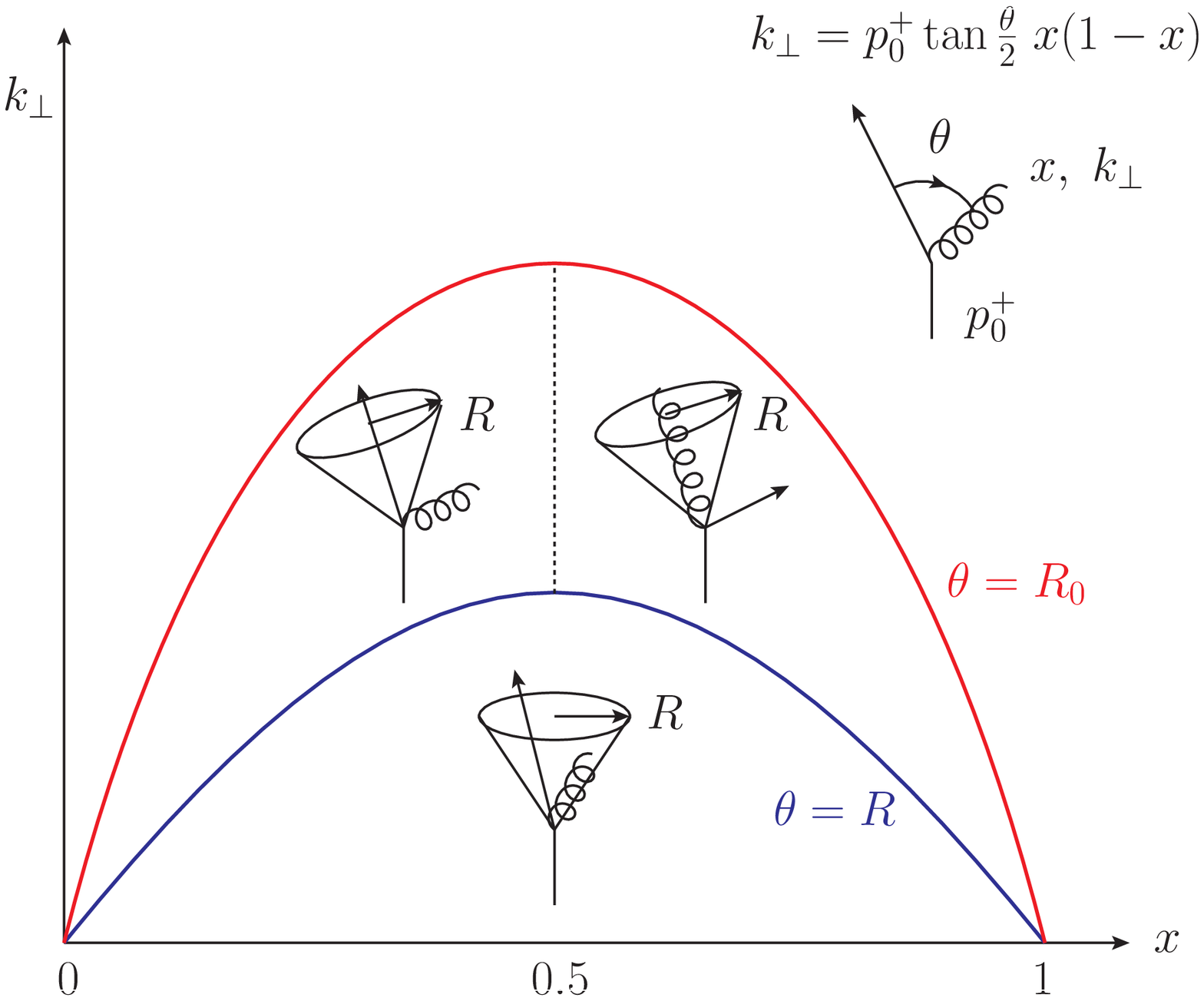}
\caption{ Illustration of the phase space regions for the calculations of the jet energy function (left panel) and the jet energy lost (right panel) at leading order. In both cases, a collinear parton splits into partons with momenta $k=(xp_0^+,k_\perp^2/xp_0^+,k_\perp)$ and $p-k$. Depending on the kinematics of the splitting, the partons which can contribute to the energy measured inside the subcone of size $r$ may change. Similarly, the partons reconstructed in the jet may also differ.}
\label{PSintegral}
\end{figure}

In fact, the collinear sector in SCET is a boosted copy of QCD \cite{Beneke:2002ph,Freedman:2011kj,Feige:2012vc,Feige:2013zla}. Realizing this, various next-to-next-to-leading order jet and beam functions were calculated bypassing the use of the more complicated collinear SCET Feynman rules \cite{Becher:2006qw,Becher:2010pd,Gaunt:2014xxa,Gaunt:2014xga,Gaunt:2014cfa}. It was also pointed out that jet and beam functions can be calculated directly from the QCD splitting functions \cite{Ritzmann:2014mka} with proper phase space ($PS$) integrations. At leading order \footnote{While splitting functions and jet energy functions are calculated at fixed order, all-order contributions from vacuum parton splittings and multiple gluon emissions are included in the resummation at next-to-leading logarithmic accuracy. However, contributions from correlated multiple scatterings with the medium, \textcolor{\highlight}{i.e. higher-order terms in the opacity expansion,} are not resummed in this work. \textcolor{\highlight}{They may affect the medium-induced splitting functions, and the impact on the jet shape modification will be investigated in the future.}}, the jet energy function associated with parton $i$ with the collinear momentum $p$ splitting into $k=(x\omega,k_\perp^2/x\omega,k_\perp)$ and $p-k$ can then be written as follows,
\be
    J_{\omega,E_r}^{i}(\mu)=\textcolor{\highlight}{\mu^{4-d}}\sum_{j,k}\int dx \textcolor{\highlight}{\frac{d k_\perp}{k^{4-d}_\perp}} {\cal P}_{i\rightarrow jk}(x, k_\perp) E_r(x, k_\perp)\;,
\ee
where ${\cal P}_{i\rightarrow jk}(x, k_\perp)$ are the collinear parton splitting functions \textcolor{\highlight}{and the $d$-dependence is suppressed}. In the presence of the medium,
\be
    {\cal P}_{i\rightarrow jk}(x, k_\perp)={\cal P}^{vac}_{i\rightarrow jk}(x, k_\perp)+{\cal P}^{med}_{i\rightarrow jk}(x, k_\perp),
\ee
which is the sum of the vacuum Altarelli-Parisi splitting functions and their medium-induced modifications \footnote{\textcolor{\highlight}{The medium-induced splitting functions in this paper were calculated using $\rm SCET_G$ at the first order in the opacity expansion.}}. $E_r(x, k_\perp)$ is the measurement function associated with the jet energy function,
\be
    E_r(x, k_\perp)={\cal M}_1+{\cal M}_2+{\cal M}_3+{\cal M}_4\;,
\ee
where
\bea
    {\cal M}_1&=&\Theta\Big(\omega x\tan\frac{r}{2}-k_\perp\Big)\Theta\Big(\omega(1-x)\tan\frac{r}{2}-k_\perp\Big)\Theta_{\rm k_T}\times p^0 \; , \\
    {\cal M}_2&=&\Theta\Big(\omega x\tan\frac{r}{2}-k_\perp\Big)\Theta\Big(k_\perp-\omega(1-x)\tan\frac{r}{2}\Big)\Theta_{\rm k_T}\times k^0 \; ,\\
    {\cal M}_3&=&\Theta\Big(k_\perp-\omega x\tan\frac{r}{2}\Big)\Theta\Big(\omega(1-x)\tan\frac{r}{2}-k_\perp\Big)\Theta_{\rm k_T}\times (p^0-k^0) \; , \\
    {\cal M}_4&=&\Theta\Big(k_\perp-\omega x\tan\frac{r}{2}\Big)\Theta\Big(k_\perp-\omega(1-x)\tan\frac{r}{2}\Big)\Theta_{\rm k_T}\times 0\;,
\eea
are the cases in which each particle is either inside the subcone of size $r$ or not. $\Theta_{\rm k_T}$ is the phase space constraint corresponding to the reconstruction of jets of size $R$ using the anti-$\rm k_T$ algorithm,
\be
    \Theta_{\rm k_T}=\Theta\Big(\omega x(1-x)\tan\frac{R}{2}-k_\perp\Big)\;.
\ee
The different phase space regions in the jet energy function calculation are illustrated in the left panel in Fig.\ref{PSintegral}. It is then straightforward to calculate the medium modification of the jet energy function,
\be
    J_{\omega,E_r}(\mu) = J^{vac}_{\omega, E_r}(\mu)+J^{med}_{\omega, E_r}(\mu).
\ee
Note that, 
%due to the LPM effect the medium modification contributes as a power correction 
due to the LPM effect the medium modification in this treatment contributes as a power correction 
because ${\cal P}^{med}_{i\rightarrow jk}(x, k_\perp)$ is finite as $k_\perp\rightarrow 0$. The renormalization group evolution of the jet energy function is then the same as in vacuum,
\be
    \frac{d J^{i}_{\omega,E_r}(\mu)}{d\ln\mu}
    =\left[-C_i\Gamma_{\rm cusp}(\alpha_s)\ln\frac{\omega^2\tan^2\frac{R}{2}}{\mu^2}-2\gamma^{i}(\alpha_s)\right]J^{i}_{\omega,E_r}(\mu),
\ee
where $i=q, g$ with $C_q=C_F$ and $C_g=C_A$ the Casimir operators of the fundamental and adjoint representations in QCD. The anomalous dimensions $\Gamma_{\rm cusp}$ and $\gamma^i$ are given in \cite{Chien:2014nsa}, and the solutions to the renormalization group evolutions have the same form,
\be
    J^{i}_{\omega,E_r}(\mu)=J^{i}_{\omega,E_r}(\mu_{j_r})\exp\left[-2C_iS(\mu_{j_r},\mu)+2A_{i}(\mu_{j_r},\mu)\right]
       \left(\frac{\omega^2\tan^2\frac{R}{2}}{\mu_{j_r}^2}\right)^{C_iA_\Gamma(\mu_{j_r},\mu)}\;,
\ee
which describe how jet energy functions evolve from the jet energy scale $\mu_{j_r}$ to arbitrary renormalization scale $\mu$. Here $S(\nu,\mu)$, $A_i(\nu,\mu)$ and $A_\Gamma(\nu,\mu)$ are the renormalization-group evolution kernels in SCET and the explicit expressions can be found in \cite{Chien:2014nsa}. The integral jet shape becomes
\be
    \Psi^i_\omega(r)
    =\frac{J^{i}_{\omega,E_r}(\mu_{j_r})}{J^{i}_{\omega,E_R}(\mu_{j_R})}\exp[-2C_iS(\mu_{j_r},\mu_{j_R})+2A_{i}(\mu_{j_r},\mu_{j_R})]
    \left(\frac{\mu^2_{j_r}}{\omega^2\tan^2\frac{R}{2}}\right)^{C_iA_\Gamma(\mu_{j_R},\mu_{j_r})}\;.
\ee
As discussed in \cite{Chien:2014nsa}, the same choice of jet energy scales
\be
    \mu_{j_r}=\omega\tan\frac{r}{2}\approx E_J\times r\;,~~~~~\mu_{j_R}=\omega\tan\frac{R}{2}\approx E_J\times R\;,
\ee
where $E_J$ is the jet energy, eliminates the large logarithms in the fixed order calculation of $J^{i}_{\omega,E_r}(\mu_{j_r})$ and $J^{i}_{\omega,E_R}(\mu_{j_R})$ because the medium modification does not introduce extra logarithms. The renormalization group evolution between $\mu_{j_r}$ and $\mu_{j_R}$ then resum the logarithms $\ln \mu_{j_r}/\mu_{j_R}\approx\ln r/R$ caused by the vacuum parton splitting.

With the resummation and the medium modification performed, the integral jet shape is the average over the jet production cross section.
\be
    \Psi(r)=\frac{1}{\sigma_{\rm total}}\sum_{i=q,g}\int_{PS} d\eta dp_T \frac{d\sigma^{i}}{d\eta dp_T}\Psi^i_\omega(r)\;,
\ee
with the proper phase space cuts ($PS$) on the jet $p_T$ and $\eta$. With the necessity, we now turn to the calculation of the jet cross section in the medium.

The jet cross section in nucleus collisions is calculated through the following expression with the Glauber modeling \cite{Miller:2007ri},
\be
    \frac{1}{\langle N_{\rm bin}\rangle}\frac{d\sigma^k_{\rm CNM}}{d\eta dp_T}
    = \sum_{ijX}\int dx_1dx_2 f_i^A(x_1;\mu_{\textcolor{\highlightA}{\rm CNM}})f_j^A(x_2;\mu_{\textcolor{\highlightA}{\rm CNM}})\frac{d\sigma_{ij\rightarrow kX}}{dx_1dx_2d\eta dp_T},
\ee
where $f^{A}_i(x;\mu_{\textcolor{\highlightA}{\rm CNM}})$ are the effective parton distribution functions (PDFs) in the nucleus $A$, and $\mu_{\textcolor{\highlightA}{\rm CNM}}$ as an argument in parton distribution functions parameterizes the cold nuclear matter effect we implement, not to be confused with the factorization scale \footnote{In this work we do not incorporate the intrinsic modifications of the parton distributions which are conventionally referred to as nuclear parton distribution functions (nPDFs). We only examine and refer to the cold nuclear matter effect from initial state parton energy loss \cite{Vitev:2007ve}. This is caused by the radiation induced in the presence of the nucleons in the nucleus by the RMS momentum exchange at the scale $\mu_{\textcolor{\highlightA}{\rm CNM}}$ with the colliding partons.}. The induced radiation of the initial-state and final-state partons can both cause a modification of the jet cross section. The former contributes as one of the cold nuclear matter (CNM) effects because the hot, dense medium is produced after the collision. It can give significant contributions \cite{Kang:2015mta} to the attenuation of the jet production cross section in central d+Au collisions at RHIC~\cite{Adare:2015gla} and p+Pb collisions at the LHC~\cite{ATLAS:2014cpa}. Although here we mainly discuss the production cross section of jets, measurements of the inclusive particle production in p+Pb collisions~\cite{Kang:2012kc} and the quenching of hadrons with very high transverse momentum in Pb+Pb collisions at the LHC~\cite{Chien:2015vja} are consistent with small CNM effects.

The calculation of the CNM effect in this paper follows \cite{Vitev:2007ve} in which the effective field theory techniques have not been applied yet. We implement the initial-state CNM effect as follows,
%\begin{eqnarray}
%\int dx \, \phi_{q}(x) \cdots & \rightarrow & \int d{x} \int d\epsilon \,
%\phi_q\left(\frac{{x}}{1-\epsilon}\right)P_q(\epsilon) \;,  \\
%\int dx \, \phi_{g}(x) \cdots & \rightarrow  & \int d\tilde{x} \int d\epsilon \,
%\phi_g\left(\frac{{x}}{1-\epsilon}\right)P_g(\epsilon) \; .
%\end{eqnarray}
\begin{eqnarray}
&&f^A_{q}(x;\mu_{\textcolor{\highlightA}{\rm CNM}})=\int d\epsilon \,
f^p_q\left(\frac{{x}}{1-\epsilon}\right)P_q(\epsilon;\mu_{\textcolor{\highlightA}{\rm CNM}}) \;,  \\
&&f^A_{g}(x;\mu_{\textcolor{\highlightA}{\rm CNM}})=\int d\epsilon \,
f^p_g\left(\frac{{x}}{1-\epsilon}\right)P_g(\epsilon;\mu_{\textcolor{\highlightA}{\rm CNM}}) \;,
\end{eqnarray}
where $f^{p}_i(x)$ are the parton distribution functions of the proton \footnote{Again we do not incorporate the nuclear parton distribution functions (nPDFs). As one example we focus on effects from initial state parton energy loss on jet shapes and cross sections. Distinguishing and combining various initial state effects in theoretical calculations, as well as detailed studies of \cite{ATLAS:2014cpa,Chatrchyan:2014hqa} where strong constraints on initial state effects can be made, are left for future work.}. $P_{q,g}(\epsilon;\mu_{\textcolor{\highlightA}{\rm CNM}})$ are the probability density for the initial-state quarks or gluons to lose a fraction $\epsilon$ of their energies due to multiple soft gluon emissions as they propagate through a large nucleus. Note that in this case the modifications of the effective PDF of the nucleus moving in the positive lightcone direction are caused by the interaction with the nucleus moving in the negative lightcone direction. The scale $\mu_{\textcolor{\highlightA}{\rm CNM}}$, which will later show up in the Results section, is the scale of the RMS momentum exchange between the colliding partons and the nucleus (not to be mixed with the factorization scale in the PDFs). The parameter controls the strength of the CNM effect implemented in the calculations.

The CNM effect can essentially be organized as the modification factors associated with the parton distribution functions,
\be
    f_i^A(x;\mu_{\textcolor{\highlightA}{\rm CNM}})=f_i^p(x)\times R_i(x;\mu_{\textcolor{\highlightA}{\rm CNM}})\;.
\ee
Note that, because the PDFs typically die off as $x\rightarrow 1$, the factors $R_i$ are smaller than 1 in this region due to the energy loss of initial state partons. This can cause partly the attenuation of the jet cross section in nucleus collisions.

Similarly, because the jet cross section dies off as the jet transverse momentum $p_T$ increases, the jet energy loss due to the final-state, medium-induced radiation will also cause the suppression of the cross section. The jet looses on average a fraction $\epsilon$ of its transverse momentum when it passes through the medium,
\be
    p_T^{med}=p_T^{vac}\times(1-\epsilon)\textcolor{\highlight}{\approx p_T^{vac}e^{-\epsilon}},
\ee
and the jet cross section in heavy ion collisions with the creation of the medium can be related to the jet cross section in heavy ion collisions without the creation of the medium as follows,
\be
    \frac{1}{\langle N_{\rm bin}\rangle}\frac{d\sigma^{k}_{med}}{d\eta dp_T}\Bigg|_{p_T}
    = \frac{1}{\langle N_{\rm bin}\rangle}\frac{d\sigma^{k}_{\rm CNM}}{d\eta dp_T}\Bigg|_{\frac{p_T}{1-\epsilon_k}}\frac{1}{1-\epsilon_k}
    \textcolor{\highlight}{\approx \frac{1}{\langle N_{\rm bin}\rangle}\frac{d\sigma^{k}_{\rm CNM}}{d\eta dp_T}\Bigg|_{p_T e^{\epsilon_k}}e^{\epsilon_k}}\;.
\ee
Here the index $k=q,~g$ labels whether the jet is quark-initiated or gluon-initiated. More precise calculations of the jet cross section in heavy ion collisions with the creation of the medium through possible factorized expressions will be investigated in the future.

The fraction $\epsilon_k$ of the jet energy loss can be calculated using the medium-induced splitting functions in $\rm SCET_G$. The same quantity is also studied and resummed in the vacuum for jets with small radii \cite{Dasgupta:2014yra} \footnote{\textcolor{\highlight}{For the jet radii used in the current experiments at the LHC (typically, $0.2 < R < 0.5$ which is not very small, and $\alpha_s\log R$ is not large), the resummation of contributions from multiple vacuum parton splittings does not have a large effect in the jet energy loss calculation (Fig.4 in \cite{Dasgupta:2014yra}). For medium-induced radiation undergoing further vacuum-like emissions, we expect a similar numerical outcome in the calculation of the medium-induced jet energy loss. Here we only include the lowest-order contribution, leaving the resummation of higher-order contributions for future work.}}. At leading-order, the collinear parton with momentum $p$ splits into partons with momenta $k$ and $p-k$. Using the anti-$\rm k_T$ algorithm \cite{Cacciari:2008gp} with radius $R$ for jet reconstruction, at this order if the angle between the final-state partons is smaller than $R$, both the particles are reconstructed in the jet and there is no lost energy. However, if the angle between the partons is larger than $R$, the more energetic parton will be reconstructed as the jet and the softer parton will be lost.

More specifically, the medium-induced energy loss of a quark jet is
\be
    \epsilon_q
    =\frac{2}{\omega}\Big[\int_0^{\frac{1}{2}} dx k^0
    +\int_{\frac{1}{2}}^1 dx (p^0-k^0)
    \Big]\int_{\omega x(1-x)\tan\frac{R}{2}}^{\omega x(1-x)\tan\frac{R_0}{2}}dk_\perp   \frac{1}{2} \Big[  {\cal P}^{med}_{q\rightarrow qg}(x,k_\perp) + {\cal P}^{med}_{q\rightarrow gq}(x,k_\perp)     \Big]    \;,
\ee
%where we have taken into account the non-singular parts  of  ${\cal P}^{med}_{q\rightarrow qg}(x,k_\perp)$  and  ${\cal P}^{med}_{q\rightarrow gq}(x,k_\perp)$ are related through $x\rightarrow 1-x$.
whereas for gluon jets,
\be
    \epsilon_g
    =\frac{2}{\omega}\Big[\int_0^{\frac{1}{2}} dx k^0
    + \int_{\frac{1}{2}}^1 dx (p^0-k^0)
    \Big]\int_{\omega x(1-x)\tan\frac{R}{2}}^{\omega x(1-x)\tan\frac{R_0}{2}}dk_\perp \frac{1}{2} \Big[
    {\cal P}^{med}_{g\rightarrow gg}(x,k_\perp)+   \sum_{q,\bar{q}}{\cal P}^{med}_{g\rightarrow q\bar q}(x,k_\perp)\Big].
\ee
Note that in the splitting $g\rightarrow gg$ the final-state partons are identical particles. Here $R$ is the angular parameter used in the jet reconstruction, and $R_0$ is of ${\cal O}(1)$ in QCD which sets the region of the use of collinear parton splitting functions.  The phase space regions in the above expressions are illustrated in the right panel of Fig.\ref{PSintegral}. Note the possibility of jet-flavor changing $q\rightarrow g$ or $g\rightarrow q$ due to parton splitting and the restricted phase space constrained by the jet algorithm. The flavor-changing probability is typically small for jets with large transverse momenta, so is the fraction of jet energy loss.

\section{Results}
\label{sec:results}

We compare our jet shape and cross section calculations to the ALICE \cite{Abelev:2013kqa}, ATLAS \cite{Aad:2012vca,Aad:2014bxa} and CMS \cite{Chatrchyan:2013kwa,CMS:prelim} measurements in lead-lead collisions with nucleon-nucleon center of mass energy at $\sqrt{s_{\rm NN}}=2.76$ TeV. We will also make predictions for the anticipated measurements of both inclusive and photon-tagged jets at the $\sqrt{s_{\rm NN}}\approx 5.1$ TeV LHC Run II. Even though the production cross section of jets associated with prompt photons is several orders of magnitude smaller than the inclusive jet cross section, photon+jet events give a clean probe of the medium because the prompt photon can be used as a robust reference to study the quenching of the recoiling jet \cite{Chatrchyan:2012gt}. Also, we will see that the shape of photon-tagged jets will serve as a new observable to amplify the differences between the vacuum and the medium parton showers due to the narrow energy profile of predominantly quark jets that recoil against the prompt photon. Measurements of jets recoiling against a weak boson ($W$/$Z$) will also be interesting to perform.

Jets were reconstructed using the anti-$\rm k_T$ algorithm \cite{Cacciari:2008gp} with $R=0.2$, 0.3, 0.4 and 0.5 in the measurements of inclusive jet cross sections. For the jet shape, the CMS collaboration used $R=0.3$. Future measurements are planned be performed with multiple jet radii, so in the jet shape calculation we include both $R=0.3$ and $0.5$ to investigate the jet radius dependence.

The inclusive jet cross sections are examined in various jet transverse momentum ($p_T^{\rm jet}$) and pseudo-rapidity ($\eta^{\rm jet}$) bins. To compare with the CMS jet shape measurement, we impose the same cuts as follows,
\be
    p_T^{\rm jet} > 100~{\rm GeV}\;,~0.3<|\eta^{\rm jet}|<2\;.
\ee
The region $|\eta^{\rm jet}|<0.3$ is excluded because of the techniques used in the background subtraction. In this paper the calculations for photon+jets events are inclusive, in the sense that we have allowed all possible photon kinematics consistent with the cuts on $p_T^{\rm jet}$ and $\eta^{\rm jet}$. Future experiments may impose cuts on the photon kinematics as well as some isolation criteria, and they will have to be implemented in the theory calculation.

The differential jet shapes measured by CMS are constructed from the transverse momenta of the charged particles with $p_T^{\rm track}>1$ GeV,
\be
    \frac{\Delta \Psi(r)}{\Delta r}=\frac{1}{N_J}\sum_{J=1}^{N_J}\frac{\Psi_J^{\rm track}(r+\delta r/2)-\Psi_J^{\rm track}(r-\delta r/2)}{\delta r}\;,
\ee
and the jet cone is divided into six annuli between $0<r<0.3$ with $\delta r=0.05$. For the calculation with $R=0.5$, we divide the jet cone into ten annuli between $0<r<0.5$ with the same $\delta r$.

It should be noted that, the ATLAS measurements used calorimeter towers with $\Delta\eta\times\Delta\phi=0.1\times0.1$ for jet reconstructions, where the detector granularity may start to play a role for the reconstruction of $R=0.2$ jets. Also, ALICE measured charged jets and CMS measured the jet shape using charged particles. The above details are not taken into account in the theory calculation yet, however, the difference will be examined in the future as the theory and the experiments progress.

For the cross section calculations of both inclusive and photon-tagged jets, we use the CTEQ5M parton distribution functions (PDFs) \cite{Tung:2006tb} and the leading-order ${\cal O}(\alpha_s^2)$ and ${\cal O}(\alpha_s\alpha_{em})$ QCD partonic cross section results. The cross sections in Pb+Pb collisions are calculated by implementing the initial-state cold nuclear matter effects and the final-state jet energy loss caused by the medium-induced splitting. To examine the CNM effect and its theoretical uncertainty, we consider three different cases in \cite{Vitev:2007ve} with the characteristic RMS medium momentum exchange scales $\mu_{\textcolor{\highlightA}{\rm CNM}}=0$ GeV, $\mu_{\textcolor{\highlightA}{\rm CNM}}=0.18$ GeV and $\mu_{\textcolor{\highlightA}{\rm CNM}}=0.35$ GeV
%corresponding to zero, small and moderate effects of the initial-state energy loss. 
corresponding to different sizes of initial-state energy loss effects.
For the consistency of the theory calculations of the jet-medium interaction, we use the same splitting kernels which were used in the description of inclusive particle production and the detailed expressions can be found in \cite{Chien:2015vja}. The coupling $g$ \footnote{$\alpha_s=g^2/4\pi$ is the strong coupling constant associated with the jet-medium interaction. We estimate the uncertainty by varying the interaction strength with fixed $g$'s within $g=2.0\pm 0.2$. The detailed medium effects absorbed in the coupling $g$ will be investigated in the future.} between the jet and the medium is typically chosen to be around $g=2.0$, and we estimate the theory uncertainty using the window $g=2.0\pm 0.2$ in most of the cases.

\textcolor{\highlight}{
As we will see below, varying the medium momentum exchange scale $\mu_{\textcolor{\highlightA}{\rm CNM}}$ between 0 GeV and 0.35 GeV 
%has a large effect on the nuclear modification factor $R_{AA}$ (defined in Eq.~(\ref{RAAjet})). 
has a large effect on the nuclear modification factor $R_{AA}$ (defined in Eq.~(\ref{RAAjet})) at high $p_T$. 
Also, an ${\cal O}(10\%)$ variation of the coupling $g$ can cause considerable change in the $R_{AA}$ values. These suggest that the inclusive jet cross section is sensitive to both initial state and final state effects. However, {\textcolor{\highlightA}{later we will see that}} the jet shape is {\textcolor{\highlightA}{less sensitive}} to the variation of $\mu_{\textcolor{\highlightA}{\rm CNM}}$ therefore allowing us to precisely probe the final state, jet-medium interaction. This in turn can allow us to disentangle and more reliably constrain the CNM effect using the $R_{AA}$ data. The typical value of $g=2.0$ used in the calculation is consistent with the jet shape modification measurement. It should also be noted that, in this paper we focus on the effect of radiative energy loss without including the effect of collisional energy loss in the jet-medium interaction. While collisional energy loss may not affect the jet shape much, it can give significant contributions to the $R_{AA}$ suppression. Detailed studies of the CNM effect and the collisional energy loss 
%will be investigated in the future.
will be performed in the future.
}

The jet shape calculated in SCET are averaged with the jet cross section to produce the final jet shape to be compared with the measurements. In the calculations of the differential jet shapes, we include the one-loop jet energy functions and their medium modifications. For the renormalization group evolution we include the two-loop cusp anomalous dimensions ($\Gamma_0$ and $\Gamma_1$) and the one-loop anomalous dimensions ($\gamma_0^{q,g}$) of the jet energy functions, as well as the two-loop running of the strong coupling constant with $\alpha_s(m_Z)=0.1172$ \cite{Becher:2008cf}. This allows us to resum the jet shape at next-to-leading logarithmic accuracy (NLL). The theoretical uncertainty will be estimated conventionally by varying the jet energy scales within a factor of 2.

\subsection{Jet shapes and cross sections in Pb+Pb collisions at $\sqrt{s_{\rm NN}} = 2.76$~TeV}
\label{sec:2760}

\begin{figure}[!t!t]
\centering
\includegraphics[width=0.49\textwidth]{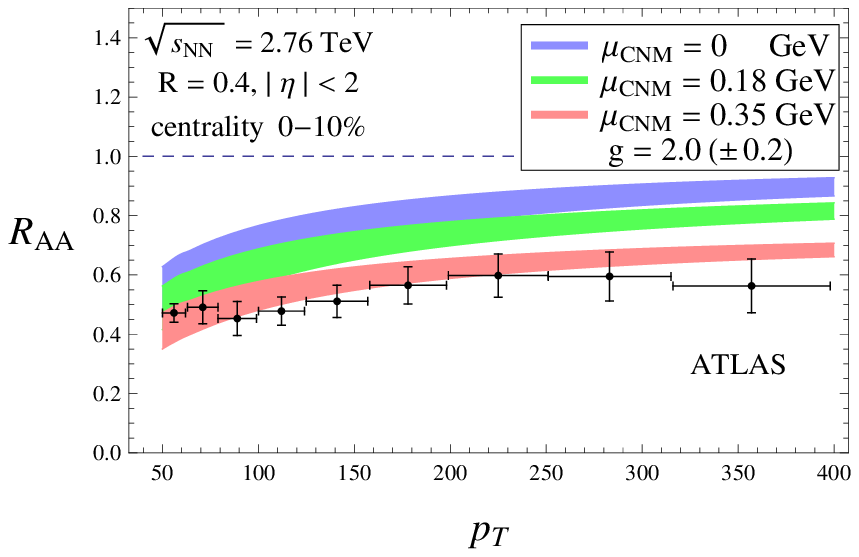}
\includegraphics[width=0.49\textwidth]{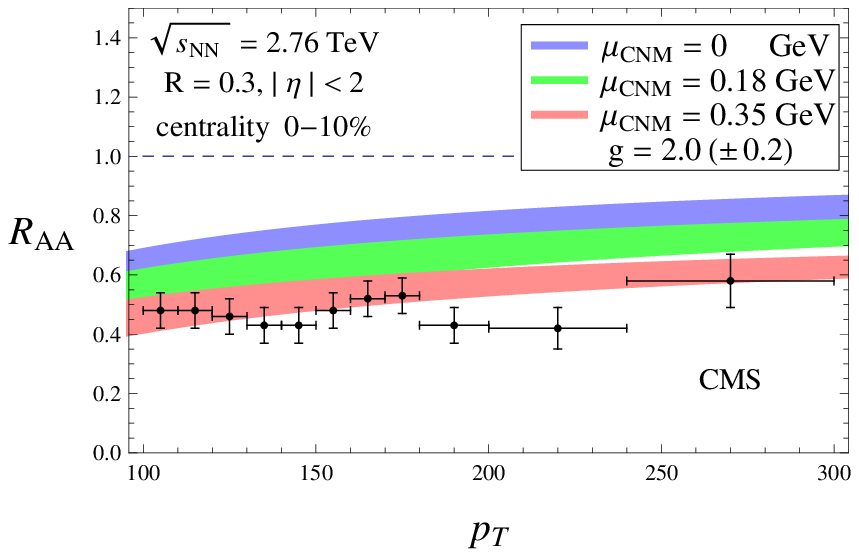}
\caption{Comparison of theoretical calculations for the nuclear modification factor $R_{AA}$ of inclusive jets as a function of the jet transverse momentum to experimental data in central Pb+Pb collisions at $\sqrt{s_{\rm NN}}=2.76$~TeV at the LHC. Bands correspond to the theoretical uncertainty estimated by varying the coupling between the jet and the medium ($g=2.0 \pm 0.2$). 
%The blue band corresponds to the calculations with no CNM effects ($\mu_{\textcolor{\highlightA}{\rm CNM}}=0$~ GeV), the green band corresponds to the calculations with small CNM effects ($\mu_{\textcolor{\highlightA}{\rm CNM}}=0.18$~GeV) and the red band corresponds to the calculations with moderate CNM effects ($\mu_{\textcolor{\highlightA}{\rm CNM}}=0.35$~GeV).
The blue band corresponds to the calculations with $\mu_{\textcolor{\highlightA}{\rm CNM}}=0$~ GeV, the green band corresponds to the calculations with $\mu_{\textcolor{\highlightA}{\rm CNM}}=0.18$~GeV and the red band corresponds to the calculations with $\mu_{\textcolor{\highlightA}{\rm CNM}}=0.35$~GeV.
Left panel: comparison to the ATLAS measurement~\cite{Aad:2014bxa} with $R=0.4$. Right panel: comparison to the CMS preliminary result~\cite{CMS:prelim} with $R=0.3$.}
\label{RAA_gCNM_ATLAS_CMS}
\end{figure}

We first present theoretical calculations and the comparisons to the measurements in Pb+Pb collisions at $\sqrt{s_{\rm NN}} = 2.76$~TeV. Fig.~\ref{RAA_gCNM_ATLAS_CMS} shows the results of the nuclear modification factor $R_{AA}$ %(defined in Eq.~(\ref{RAAjet}))
of inclusive jet productions for jets with pseudo-rapidity $|\eta| < 2$ in the 0 - 10\% centrality class as a function of jet transverse momenta $p_T$. We choose to present the calculation for jets with $p_T > 50$~GeV. For jets with small $p_T$, hadronization effects can play a significant role and our perturbative calculations will have to include these non-perturbative contributions. On the other hand, with the huge underlying event backgrounds, the reconstruction of lower $p_T$ jets in experiments becomes increasingly difficult.

\begin{figure}[!t!t]
\centering
\includegraphics[width=0.49\textwidth]{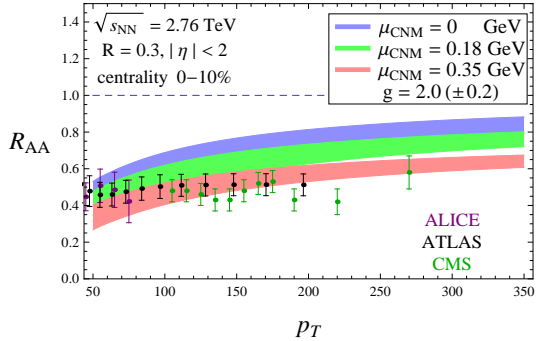}
\caption{Comparison of theoretical calculations for the suppression of inclusive jets with $R=0.3$ as a function of the jet transverse momentum to experimental data in Pb+Pb collisions at $\sqrt{s_{\rm NN}}=2.76$~TeV. The curves are the jet $R_{AA}$ calculations for central colliisons, but we have included the data from the ALICE charged jets $R_{CP}$~\cite{Abelev:2013kqa}, ATLAS calorimeter jet $R_{CP}$~\cite{Aad:2012vca} and CMS jet $R_{AA}$~\cite{CMS:prelim} measurements for comparison, assuming that the suppression in peripheral collisions is small. Shown are the same studies as in Fig.~2 of the theory sensitivity on the coupling $g$ between the jet and the QGP medium, as well as the momentum transfer $\mu_{\textcolor{\highlightA}{\rm CNM}}$ between the initial-state parton and the nucleus.
%\ref{RAA_gCNM_ATLAS_CMS}.
}
\label{AA_gCNM_ALL}
\end{figure}

Fig. \ref{RAA_gCNM_ATLAS_CMS} and \ref{AA_gCNM_ALL} present the theoretical sensitivity of $R_{AA}$ on cold nuclear matter effects caused by initial state parton energy loss. The magnitude of CNM effects depends strongly on the magnitude of the momentum transfer between the incident partons and the nucleus. 
%We show three different cases with no CNM effects (represented by the blue bands with $\mu_{\textcolor{\highlightA}{\rm CNM}}=0$~GeV), small CNM effects (represented by the green bands with $\mu_{\textcolor{\highlightA}{\rm CNM}}=0.18$~GeV), and moderate CNM effects (represented by the red bands with $\mu_{\textcolor{\highlightA}{\rm CNM}}=0.35$~GeV).
We show three different cases with $\mu_{\textcolor{\highlightA}{\rm CNM}}=0$~GeV (represented by the blue bands), $\mu_{\textcolor{\highlightA}{\rm CNM}}=0.18$~GeV (represented by the green bands), and $\mu_{\textcolor{\highlightA}{\rm CNM}}=0.35$~GeV (represented by the red bands).
Here, $\mu_{\textcolor{\highlightA}{\rm CNM}} = \sqrt{\Delta {\bf q}_T^2}$ which is the root-mean-square of the transverse momentum transfer per scattering between the incident hard parton and the constituents of the nucleus. On the other hand, the strength of the final state jet-medium interaction is controlled by the coupling $g$ between the collinear partons and the QGP medium. The width of the bands in the plots corresponds to the variation of $g=2.0 \pm 0.2$.

In the left panel of Fig.~\ref{RAA_gCNM_ATLAS_CMS}, we present the comparison of the theoretical results to the ATLAS $R_{AA}$ measurements with $R=0.4$.  We find that the final state jet-medium interaction alone is not sufficient to describe the suppression of the cross section at high transverse momenta, 
%and the inclusion of small CNM effects provides a good description of the experimental data (green band). Larger CNM effects result in the amount of cross section suppression not compatible with the data.
and the inclusion of CNM effects with $\mu_{\textcolor{\highlightA}{\rm CNM}}=0.35$~GeV provides a good description of the experimental data (red band). 
The comparison of our calculations to the preliminary CMS data with $R=0.3$ is shown in the right panel of Fig.~\ref{RAA_gCNM_ATLAS_CMS} with similar conclusions.

In Fig.~\ref{AA_gCNM_ALL}, we include a comparison to the ALICE $R_{CP}$ measurement of the suppression of $R=0.3$ charged jets~\cite{Abelev:2013kqa}, i.e. jets reconstructed with the charged particle tracks. 
%We find that the calculation with small CNM effects are consistent with the quenching of charged jets in the comparatively low $p_T$ region. 
We find that the calculation with $\mu_{\textcolor{\highlightA}{\rm CNM}}=0.35$~GeV is consistent with the quenching of charged jets in the comparatively low $p_T$ region.
In the high intra-jet particle multiplicity limit, from isospin symmetry we expect that charged jets will exhibit similar characteristics as the full reconstructed calorimeter jets, therefore with similar cross section suppression. A comparison to the ATLAS $R_{CP}$ measurement with $R=0.3$ is also presented in the same plot.

\begin{figure}[!t!t]
\centering
\includegraphics[width=0.49\textwidth]{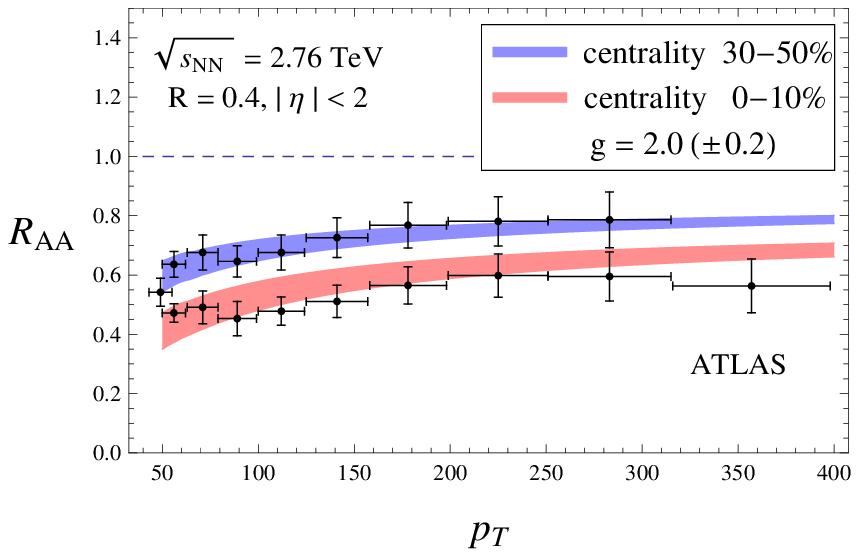}
\includegraphics[width=0.49\textwidth]{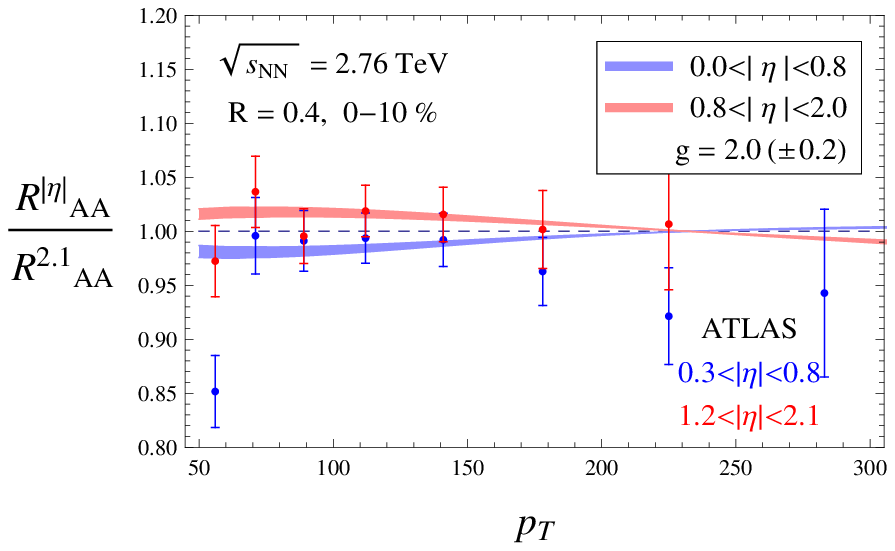}
\caption{Left panel: comparison of theoretical calculations for the nuclear modification factor $R_{AA}$ of inclusive jets as a function of the jet transverse momentum to experimental data in $\sqrt{s_{\rm NN}}=2.76$~TeV Pb+Pb collisions at the LHC, with different collision centralities. Two centrality classes, 0 - 10\%  and 30 - 40\%, are considered. The bands corresponds to the variation of the coupling $g=2\pm 0.2$ between the jet and the medium in the calculations, 
%and small CNM effects ($\mu_{\textcolor{\highlightA}{\rm CNM}}=0.18$~GeV) are implemented. 
and CNM effects with $\mu_{\textcolor{\highlightA}{\rm CNM}}=0.35$~GeV are implemented. 
The data is from ATLAS~\cite{Aad:2014bxa} with $R=0.4$. Right panel: comparison of theoretical calculations for the ratios of jet $R_{AA}$'s within different pseudo-rapidity bins as a function of the jet transverse momentum to the ATLAS experimental data~\cite{Aad:2014bxa} in central Pb+Pb collisions. The blue band corresponds to the ratio $R_{AA}(0.0<|\eta|<0.8)/R_{AA}(0.0<|\eta|<2.0)$, and the red band corresponds to the ratio $R_{AA}(0.8<|\eta|<2.0)/R_{AA}(0.0<|\eta|<2.0)$. The derived data corresponds to $R_{AA}(0.3<|\eta|<0.8)/R_{AA}(0.0<|\eta|<2.1)$ and $R_{AA}(1.2<|\eta|<2.1)/R_{AA}(0.0<|\eta|<2.1)$.}
\label{RAA_cent_rap}
\end{figure}

Next, we examine the centrality and the rapidity dependence of inclusive jet suppressions in Pb+Pb collisions at
$\sqrt{s_{\rm NN}}=2.76$~TeV. In the left panel of Fig.~\ref{RAA_cent_rap}, we show the theoretical calculations
of the inclusive jet $R_{AA}$ in central and mid-peripheral collisions and the comparison to the ATLAS measurements in
the 0 - 10\% and 30 - 40\% centrality classes. The framework for the evaluation of the medium-induced splitting
functions and their implementation in the calculation of the jet cross sections in heavy ion collisions
capture the relative contributions of the vacuum and the in-medium parton showers as a function of the size and the
density/temperature of the QGP, which is controlled by the collision centrality. The attenuation of jet production in
peripheral collisions is smaller than the attenuation in central collisions, which is consistent
with the one seen in the quenching of inclusive hadrons~\cite{Chien:2015vja}.  The right panel illustrates the dependence of jet suppression on the pseudo-rapidity $\eta$ of jets. Shown is the $R_{AA}$ of inclusive jets as a function of $p_T$ in the $0<|\eta|<0.8$ and $0.8<|\eta|<2.0$ intervals relative to the $R_{AA}$ in the $0<|\eta|<2.1$ interval. Note that within experimental uncertainties these ratios are consistent with unity. Few percent differences in the theoretical calculations capture the qualitative feature in the ATLAS data.

\begin{figure}[!t!t]
\centering
\includegraphics[width=0.49\textwidth]{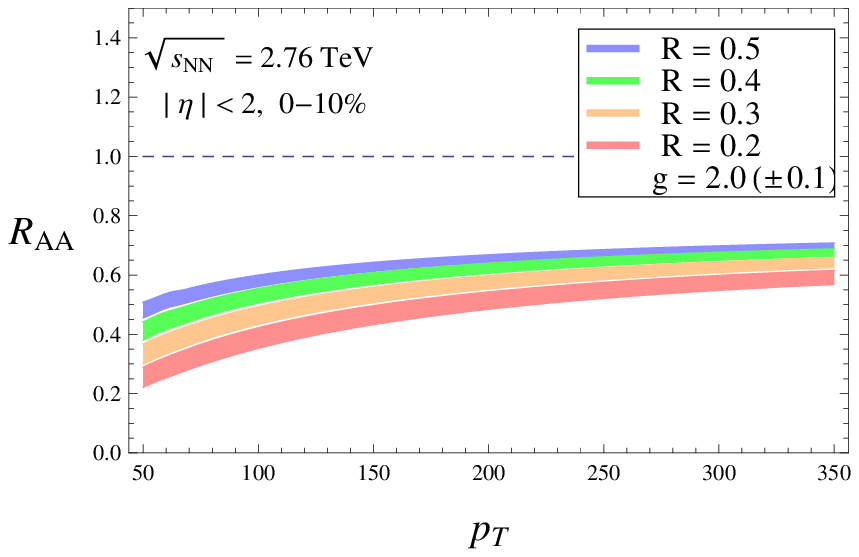}
\includegraphics[width=0.49\textwidth]{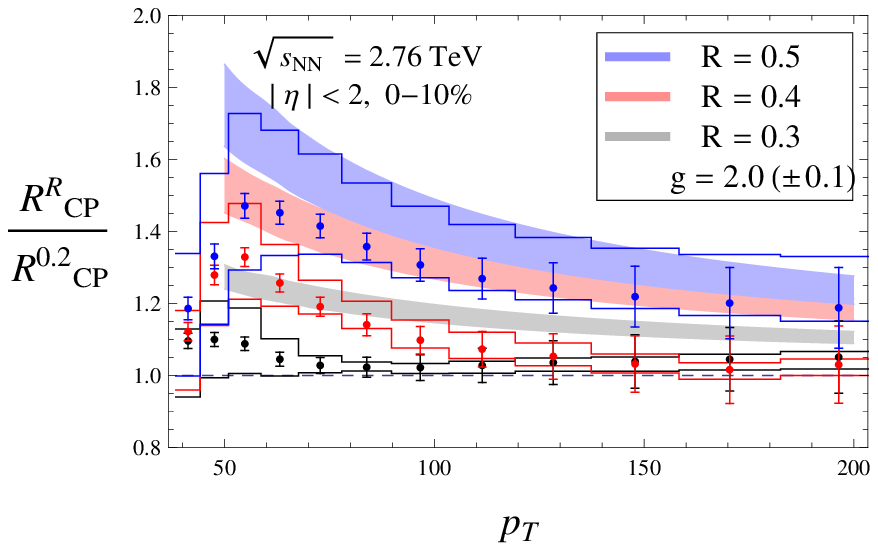}
\caption{Left panel: theoretical calculations for the nuclear modification factor $R_{AA}$ of inclusive jets with $R=0.2$, 0.3, 0.4 and 0.5 as a function of the jet transverse momentum in central Pb+Pb collisions at $\sqrt{s_{\rm NN}}=2.76$~TeV at the LHC. 
%The calculations implement small CNM effects, 
The calculations implement CNM effects with $\mu_{\textcolor{\highlightA}{\rm CNM}}=0.35$~GeV, 
and a reduced range of the coupling between the jet and the medium ($g=2\pm 0.1$) is used for estimating theoretical uncertainties. This way the bands do not overlap much and the jet-radius dependence of jet quenching can be better illustrated. Right panel: comparison of theoretical calculations for the central-to-peripheral $R_{CP}$ ratios as a function of the jet transverse momentum for inclusive jets with different jet radii to data in central Pb+Pb collisions at $\sqrt{s_{\rm NN}}=2.76$~TeV at the LHC. The bands and the data correspond to $R_{CP}(R=0.3,~0.4,~0.5)/R_{CP}(R=0.2)$. The data is from ATLAS~\cite{Aad:2012vca}.}
\label{RCP_Rdep}
\end{figure}

An important check of the calculations is to examine the jet radius $R$ dependence of the jet cross section and its medium suppression. We illustrate the different suppression patterns with different jet radii in the left panel of Fig.~\ref{RCP_Rdep}. We consider central Pb+Pb collisions at the LHC for illustration. The smaller range of the coupling $g=2.0\pm 0.1$ is used to estimate the theory uncertainty for better separation of the results corresponding to different radii. The red band represents $R=0.2$, the orange band $R=0.3$, the green band $R=0.4$ and the blue band $R=0.5$. The smaller the jet radius, the larger the suppression of the cross section.
The jet radius dependence of $R_{AA}$ is indeed observed by the ATLAS collaboration~\cite{Aad:2012vca} and shown in the right panel of     Fig.~\ref{RCP_Rdep}. More specifically, the central-to-peripheral ratio $R_{CP}$ of jet cross sections as a function of $p_T$,
defined as
\begin{equation}
R_{CP}^{R}(p_T) =
{      \langle  N_{\rm bin}^{\rm per}  \rangle
    \frac{d\sigma_{AA}^{\rm cen}(p_T,R)}{d\eta d^2 p_T} } \Big/
{ \langle  N_{\rm bin}^{\rm cen}  \rangle
\frac{ d\sigma_{AA}^{\rm per}(p_T,R)}{d\eta d^2 p_T}  }
= \frac{R_{AA}^{\rm cen}(p_T,R)}{R_{AA}^{\rm per}(p_T,R)}\; .
\label{RCPjet}
\end{equation}
is compared to the one for $R=0.2$.  In Eq.~(\ref{RCPjet})
$\langle  N_{\rm bin}^{cen.}  \rangle$ and  $ \langle  N_{\rm bin}^{per.}  \rangle $ are the mean of the number
of binary nucleon-nucleon scattering in central and peripheral Pb+Pb collisions. The calculation predicts qualitatively the transverse momentum dependence of the $R_{CP}$ ratios and provides a quantitative description of this observable when the two radii are sufficiently different.
For small radii the calculation over-predicts the difference in the quenching patterns of inclusive jets. Such discrepancy may
be caused by the fact that we only evaluate the medium-induced splitting functions at the lowest non-trivial order,
or that we neglect further dissipation of the jet energy through collisional processes in the medium~\cite{Neufeld:2011yh}. The resummation of $\log R$ for small-$R$ jet cross section can also play a role \cite{Dasgupta:2014yra,Chien:2015cka}. New experimental measurements will be very useful to further examine this jet radius dependence.

The jet shape can give complimentary information of the in-medium parton shower beyond the study of the jet cross section. Since the differential jet shape is normalized by $\int_0^R \rho(r) dr = 1 $, we expect any enhancement (attenuation) at $r \approx R$ to be correlated with the attenuation (enhancement) at small/intermediate values of $r$. This behavior was qualitatively seen in the attempt to calculate the jet shape modification using the traditional energy loss approach~\cite{Vitev:2008rz}, but the theory was not able to provide quantitative description of the CMS data~\cite{Chatrchyan:2013kwa}.

\begin{figure}[!t!t]
\centering
\includegraphics[width=0.49\textwidth]{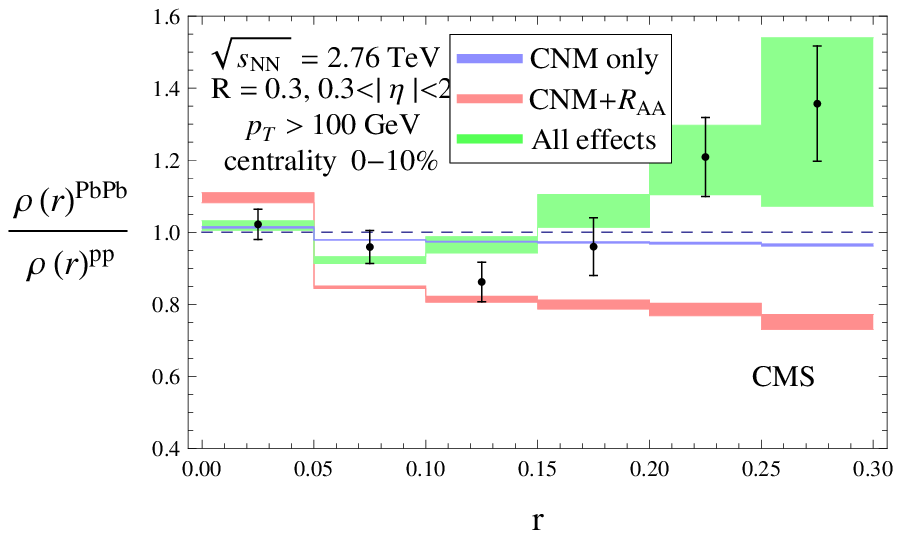}
\includegraphics[width=0.49\textwidth]{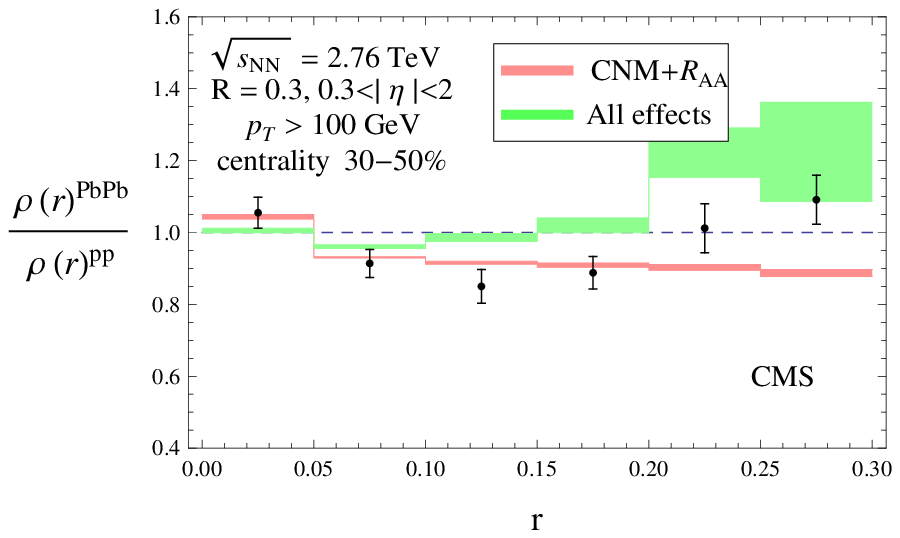}
\caption{Comparison of theoretical calculations for the modification of differential jet shapes of inclusive jets in Pb+Pb central (left panel) and peripheral (right panel) collisions at  $\sqrt{s_{\rm NN}}=2.76$~TeV at the LHC. The modification is presented as the ratio of the jet shapes $\rho^{\rm PbPb}(r)/\rho^{\rm pp}(r)$ in Pb-Pb and p-p collisions. We impose cuts on the jet transverse momentum $p_T > 100$~GeV and pseudo-rapidity $0.3<|\eta|<2.0$ of jets. The coupling between the jet and the medium is fixed at $g=2$. The blue band corresponds to the calculations including only the CNM effects, the red band adds the jet-medium interaction but with the jet-by-jet shape modification turned off, and the green band correspond to the full calculation. The theoretical uncertainty is estimated by varying the jet energy scales $\frac{1}{2}\mu_{j_R}<\mu<2\mu_{j_R}$ in the calculations. The data is from CMS~\cite{Chatrchyan:2013kwa} with $R=0.3$.}
\label{Shape_Mod_CMS}
\end{figure}

In Fig.~\ref{Shape_Mod_CMS} we compare the theoretical calculations of the jet shape modification in central and mid-peripheral Pb+Pb collisions at $\sqrt{s_{\rm NN}}=2.76$~TeV at the LHC to the CMS data~\cite{Chatrchyan:2013kwa} with $R=0.3$. In the calculations shown here, we fix the coupling between the collinear partons and the medium to be $g=2$ and estimate the theory uncertainty from the variation of the jet energy scales. {\textcolor{\highlightA}{Jet shape modifications can partly come from the change of the relative fraction of quark jets and gluon jets since quark and gluon jet shapes differ. The blue band represents the result with only the CNM effect, and the deviation from 1 is very small. This implies that the CNM effect does not significantly affect the {\sl relative} fractions of quark and gluon jets therefore the jet shape does not change much, whereas it can result in large $R_{AA}$ suppressions in the {\sl absolute} inclusive jet cross sections. We show the blue band only for central collisions to demonstrate this point, and the jet shape is not as sensitive to the CNM effect implemented in this paper as the inclusive jet cross section is.} The red band corresponds to the calculation including the cross section suppression but assuming that the jet-by-jet shape remains the same as in proton collisions. Here, the fraction of quark jets significantly increases since the cross section of gluon jets are more suppressed by the jet-medium interaction. This leads to the narrowing of the jet energy profile. The other contribution to jet shape modifications comes directly from the modification of the in-medium parton shower which leads to the broadening of jets. The green band includes all the above physics inputs which result in the attenuation at mid $r$ and the enhancement at $r\approx R$ of the jet shape. This gives, for the first time, a quantitative understanding of the jet shape modification in heavy ion reactions. The calculations for central collisions are shown in the left panel of Fig.~\ref{Shape_Mod_CMS}, and the ones for peripheral collisions are shown in the right panel.

\subsection{Theoretical predictions for jet shapes and cross sections in Pb+Pb collisions at $\sqrt{s_{\rm NN}}\approx 5.1$~TeV }
\label{sec:5100}

In this section we make predictions for the anticipated jet shape and cross section measurements at the $\sqrt{s_{\rm NN}}\approx 5.1$~TeV LHC Run II. We show the predictions for both inclusive and photon-tagged jets and compare the two. As we will see, by studying processes with different compositions of quark jets and gluon jets, we are able to examine the physical understanding of jet shapes and cross sections which allows us to use quark and gluon jets as independent probes of the QGP.

\begin{figure}[!t!t]
\centering
\includegraphics[width=0.49\textwidth]{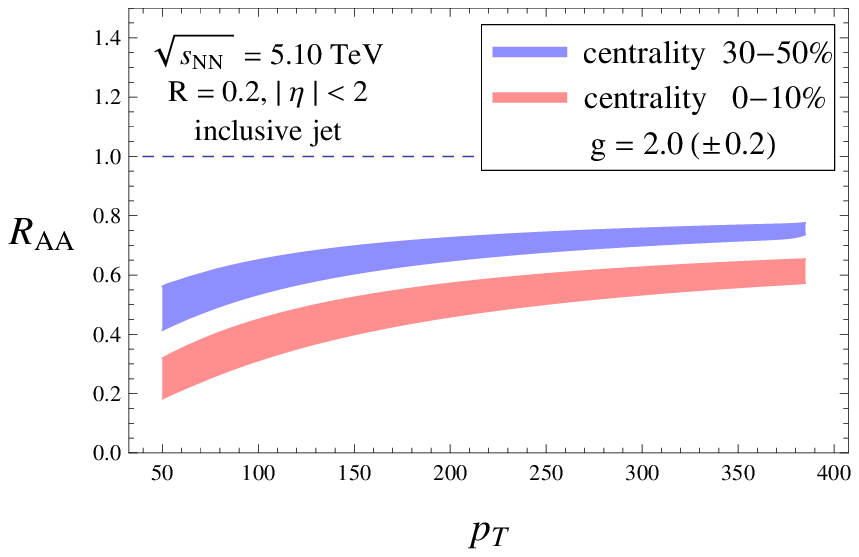}
\includegraphics[width=0.49\textwidth]{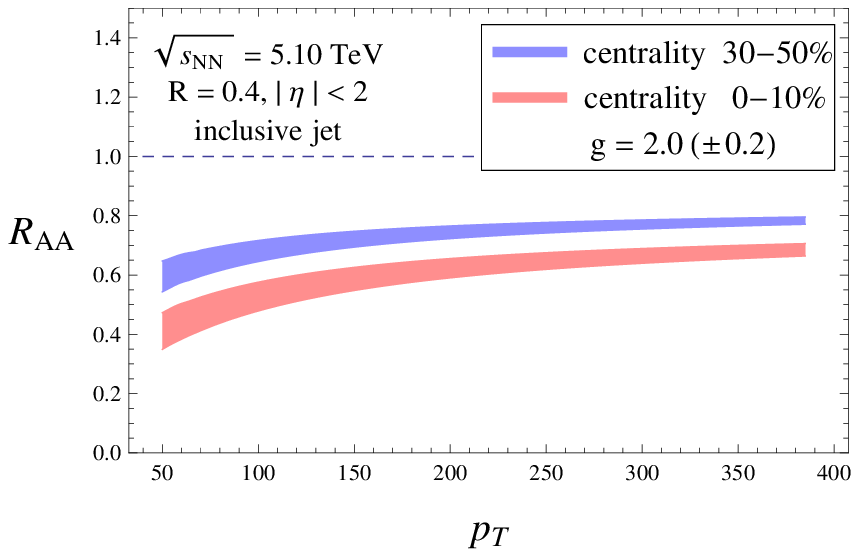}
\caption{Theoretical predictions of the nuclear modification factor $R_{AA}$ as a function of jet transverse momentum $p_T$ for inclusive jets, with $R=0.2$ (left panel) and $R=0.4$ (right panel) in Pb+Pb collisions at $\sqrt{s_{\rm NN}}\approx 5.1$ TeV at the LHC. The coupling between the jet and the medium is varied to estimate the theoretical uncertainties ($g=2.0\pm 0.2$), 
%and small CNM effects ($\mu_{\rm CNM} = 0.18$~GeV) are implemented in the calculations. 
and CNM effects with $\mu_{\rm CNM} = 0.35$~GeV are implemented in the calculations.
The red band corresponds to the collision centrality 0 - 10\% and the blue band corresponds to the collision centrality 30 - 50\%. }
\label{Predict_RAA_INCL}
\end{figure}

Fig. \ref{Predict_RAA_INCL} shows the predictions of the nuclear modification factor $R_{AA}$ of inclusive jet cross sections in central (red band, to be compared with future measurements with centrality 0 - 10\%) and mid-peripheral (blue band, to be compared with future measurements with centrality 30 - 50\%) collisions at $\sqrt{s_{\rm NN}}\approx 5.1$~TeV. The left panel corresponds to jets with $R=0.2$ and the right panel shows the $R=0.4$ results. Larger suppression is seen in central collisions and for jets with smaller radius. Again the theoretical uncertainties are estimated by varying the coupling $g=2.0\pm0.2$. 
%All the plots in this section include the implementation of small CNM effects with $\mu_{\textcolor{\highlightA}{\rm CNM}}=0.18$ GeV since the cases of $\mu_{\textcolor{\highlightA}{\rm CNM}}=0$ GeV and $\mu_{\textcolor{\highlightA}{\rm CNM}}=0.35$ GeV are not consistent with the $\sqrt{s_{\rm NN}} = 2.76$~TeV data.
All the plots in this section include the implementation of CNM effects with $\mu_{\textcolor{\highlightA}{\rm CNM}}=0.35$ GeV since the cases of $\mu_{\textcolor{\highlightA}{\rm CNM}}=0$ GeV and $\mu_{\textcolor{\highlightA}{\rm CNM}}=0.18$ GeV are not consistent with the $\sqrt{s_{\rm NN}} = 2.76$~TeV data.

\begin{figure}[!t!t]
\centering
\includegraphics[width=0.49\textwidth]{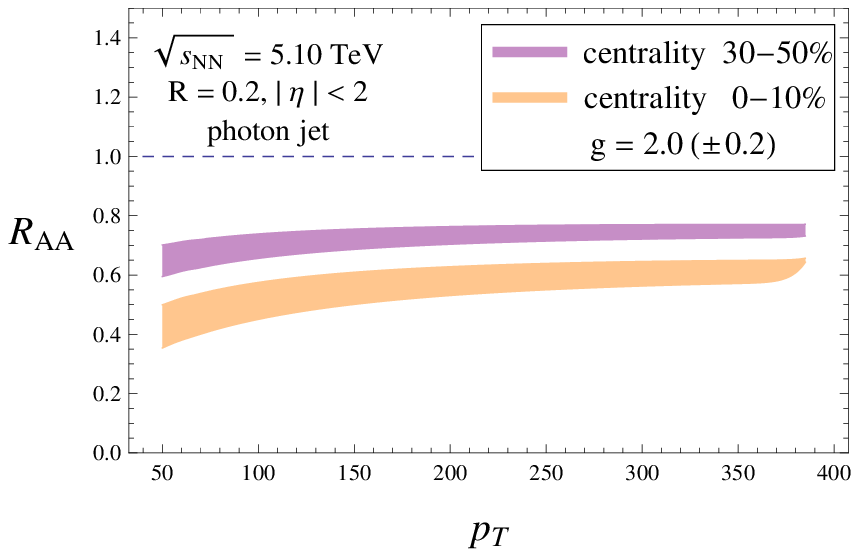}
\includegraphics[width=0.49\textwidth]{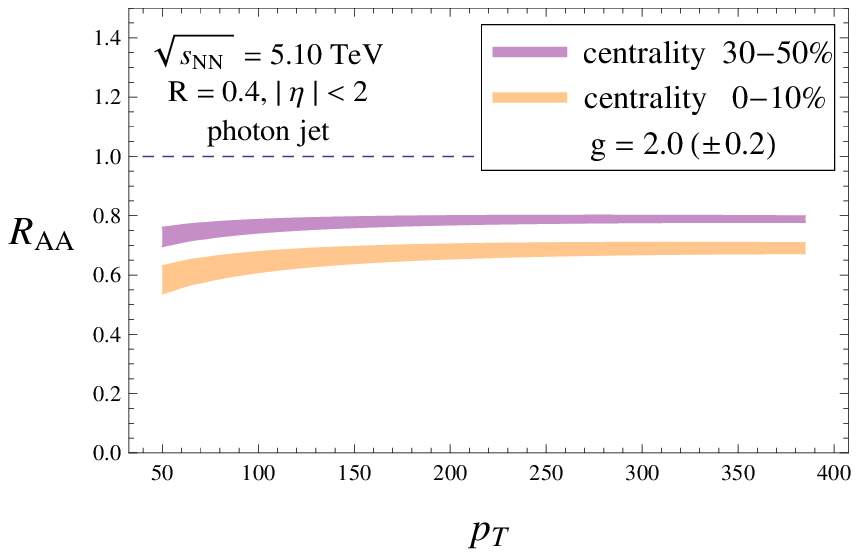}
\caption{Theoretical predictions of the nuclear modification factor $R_{AA}$ as a function of jet transverse momentum $p_T$ for jets recoiling against a prompt photon, with $R=0.2$ (left panel) and $R=0.4$ (right panel) in Pb+Pb collisions at $\sqrt{s_{\rm NN}}\approx 5.1$ TeV at the LHC. The theoretical uncertainty estimation and the implementation of CNM effects are as in Figure~7. The orange band corresponds to the collision centrality 0 - 10\% and the purple band corresponds to the collision centrality 30 - 50\%.}
\label{Predict_RAA_PHOT}
\end{figure}

Fig. \ref{Predict_RAA_PHOT} shows the predictions of the nuclear modification factor $R_{AA}$ of jet cross sections for photon+jet events in central (orange band) and mid-peripheral (purple band) collisions at $\sqrt{s_{\rm NN}}\approx 5.1$~TeV. The theoretical uncertainty is estimated with $g=2.0\pm0.2$. The jet suppression shows similar dependence on the centrality and the jet radius as the one for inclusive jets. Note that, the cross section suppression as a function of $p_T$ is more flat for photon-tagged jets compared to the one for inclusive jets. This can be seen more clearly in Fig. \ref{PREDICT_RAA_PHOTvsJET} in which we superimpose Fig. \ref{Predict_RAA_INCL} and Fig. \ref{Predict_RAA_PHOT} together. Also, for each choice of jet radius and collision centrality, the $R_{AA}$'s for inclusive jets and photon-tagged jets converge at high $p_T$. This comes from the fact that, jets with higher $p_T$ will be less modified by the interaction with the medium. 
%In the high $p_T$ limit the effect from final-state jet energy loss will decrease to zero because it is expected to be suppressed by the ratio between the medium temperature and the jet $p_T$.
In the high $p_T$ limit the effect from final-state jet energy loss will decrease to zero because it is expected to be suppressed by the LPM effect and parametrically proportional to the ratio between the medium temperature and the jet $p_T$.
In this regime the cross section suppression is mainly due to the initial-state CNM effects which we use the same implementations in this section. At lower $p_T$, the cross section suppression for photon-tagged jets is smaller than the suppression for inclusive jets. This is because the jets in photon+jet events are predominantly quark jets and the medium causes less jet energy loss as opposed to gluon jets.

\begin{figure}[!t!t]
\centering
\includegraphics[width=0.49\textwidth]{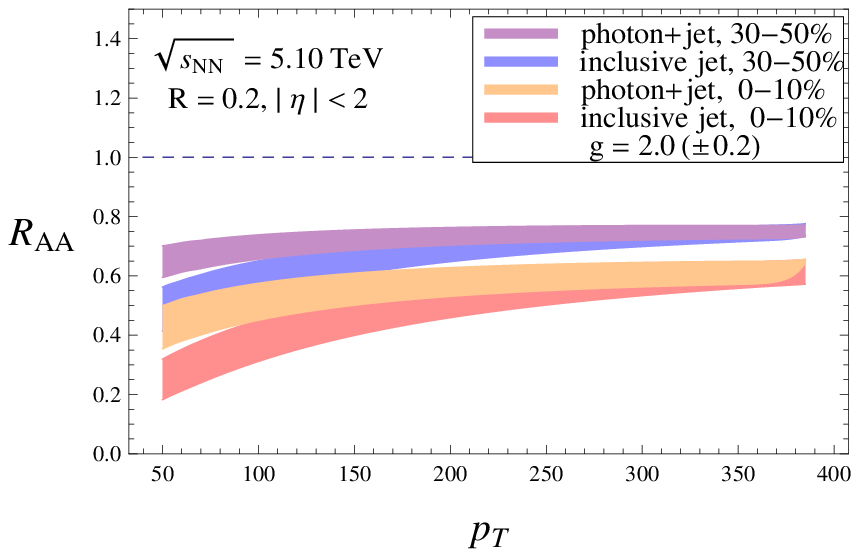}
\includegraphics[width=0.49\textwidth]{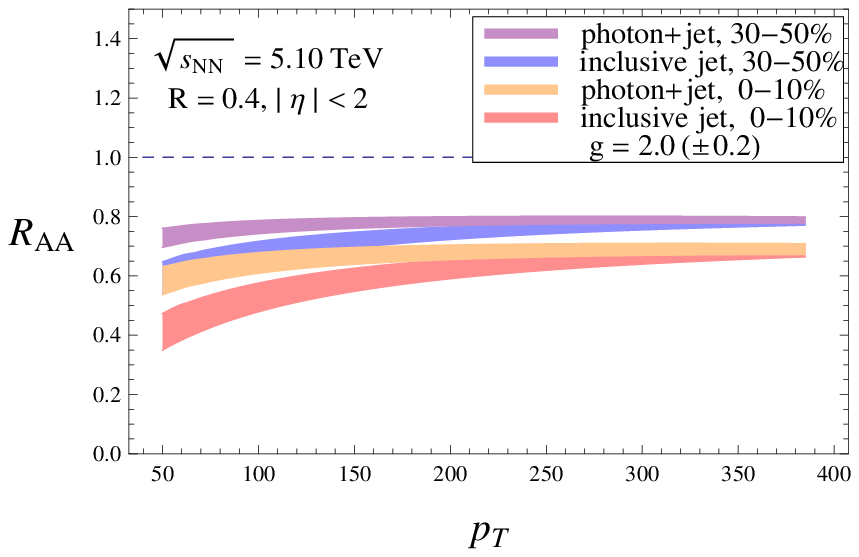}
\caption{Comparison of the theoretical predictions of the nuclear modification factor $R_{AA}$ as a function of jet transverse momentum $p_T$ for inclusive jets and photon-tagged jets, with $R=0.2$ (left panel) and $R=0.4$ (right panel) in central and mid-peripheral Pb+Pb collisions at $\sqrt{s_{\rm NN}}\approx 5.1$ TeV at the LHC. The coupling between the jet and the QCD medium is between $g=2.0\pm 0.2$, 
%and small CNM effects are implemented. 
and CNM effects with $\mu_{\rm CNM} = 0.35$ GeV are implemented.
The bands correspond to: inclusive jets, 0 - 10\% (red), inclusive jets, 30 - 50\% (blue),  photon + jet, 0 - 10\% (orange), photon + jet, 30 - 50\% (purple).}
\label{PREDICT_RAA_PHOTvsJET}
\end{figure}

We can see more clearly the difference of quark-jet and gluon-jet fractions for inclusive jets and photon-tagged jets in Fig. \ref{DIFFshape_PHOTvsINCL}. The figure shows the jet shapes for inclusive jets (blue band) and photon-tagged jets (red band) in proton collisions at $\sqrt{s}\approx 5.1$ TeV at the LHC. Here we impose the same cuts on the jet transverse momentum $p_T>100$ GeV and pseudo-rapidity $0.3<|\eta|<2.0$ as the ones in the CMS measurements at $\sqrt{s_{\rm NN}}= 2.76$ TeV. The left panel shows the jet shape for jets with $R=0.3$ and the right panel shows the one for $R=0.5$. In either case, we see that the shapes of photon-tagged jets are clearly narrower than the shapes of inclusive jets, which indicates that photon-tagged jets are predominately quark-initiated, and inclusive jets have a significant fraction of gluon-initiated ones.

\begin{figure}[!t!t]
\centering
\includegraphics[width=0.49\textwidth]{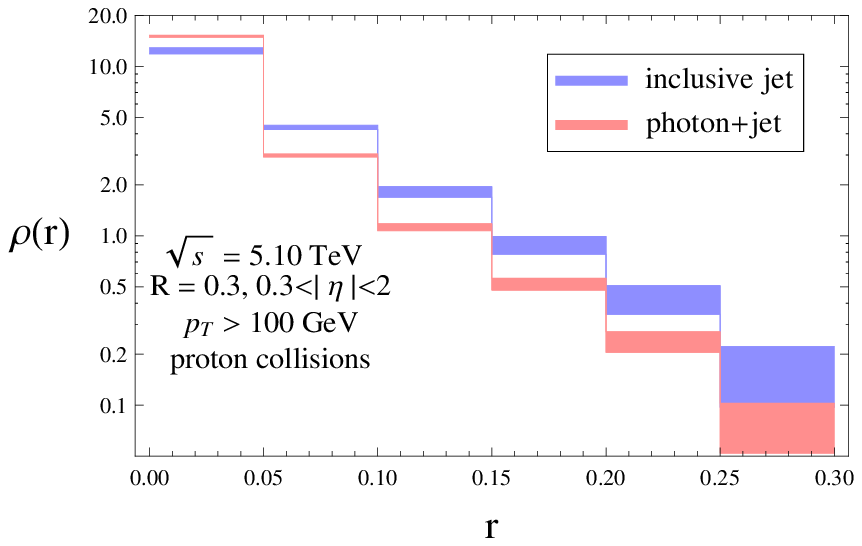}
\includegraphics[width=0.49\textwidth]{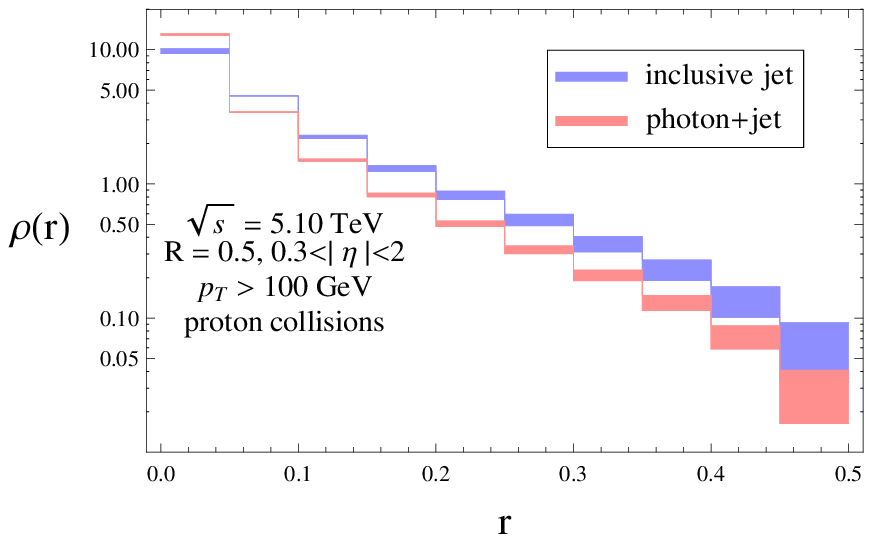}
\caption{Comparison of the theoretical predictions of the differential jet shapes for inclusive jets (blue bands) and photon-tagged jets (red bands), with $R=0.3$ (left panel) and $R=0.5$ (right panel) in $\sqrt{s} \approx 5.1$~TeV p+p collisions, with next-to-leading logarithmic accuracy. We impose cuts on the jet transverse momentum $p_T > 100$~GeV and pseudo-rapidity $0.3 < |\eta| < 2$. The bands correspond to the theoretical uncertainties estimated by varying the jet energy scales between $\frac{1}{2}\mu_{j_R}<\mu<2\mu_{j_R}$ in the calculations. Note the narrower energy profile of photon-tagged jets which are mostly quark-initiated.}
\label{DIFFshape_PHOTvsINCL}
\end{figure}

With the predictions for the baseline jet shapes in proton collisions at $\sqrt{s}\approx 5.1$ TeV, finally we now present the predictions for the modification of jet shapes in Pb+Pb collisions at $\sqrt{s_{\rm NN}}\approx 5.1$~TeV. Fig. \ref{RdiffSHAPE_INCLvsPHOT_R0.3} shows the jet shape modification for both inclusive jets (green band) and photon-tagged jets (blue band) with $R=0.3$ in central (left panel) and mid-peripheral (right panel) collisions. The theoretical uncertainty is again estimated by varying the jet energy scales in the SCET calculations. 
%While the modification of inclusive jet shapes at $\sqrt{s_{\rm NN}}\approx 5.10$~TeV 
While the modification of inclusive jet shapes at $\sqrt{s_{\rm NN}}\approx 5.1$~TeV 
is similar to the modification at $\sqrt{s_{\rm NN}}= 2.76$~TeV, we see that the modification of photon-tagged jet shapes has a different pattern, which shows clearly the broadening of jets. This is consistent with the fact that photon-tagged jets are predominately quark-initiated so that different suppressions of quark and gluon jet cross sections in the medium do not modify the jet shape as much: the gluon-jet contribution to the photon-tagged jet shape is not much.

Fig. \ref{RdiffSHAPE_INCLvsPHOT_R0.5} shows similarly the predictions for the jet shape modification of inclusive jets and photon-tagged jets with $R=0.5$ at the $\sqrt{s_{\rm NN}}\approx 5.1$~TeV LHC. The features discussed in Fig. \ref{RdiffSHAPE_INCLvsPHOT_R0.3} remain the same. However, because of the larger jet radius the jet cross section suppression is smaller for $R=0.5$ jets. This leads to the clearer effect of jet broadening.

\begin{figure}[!t!t]
\centering
\includegraphics[width=0.49\textwidth]{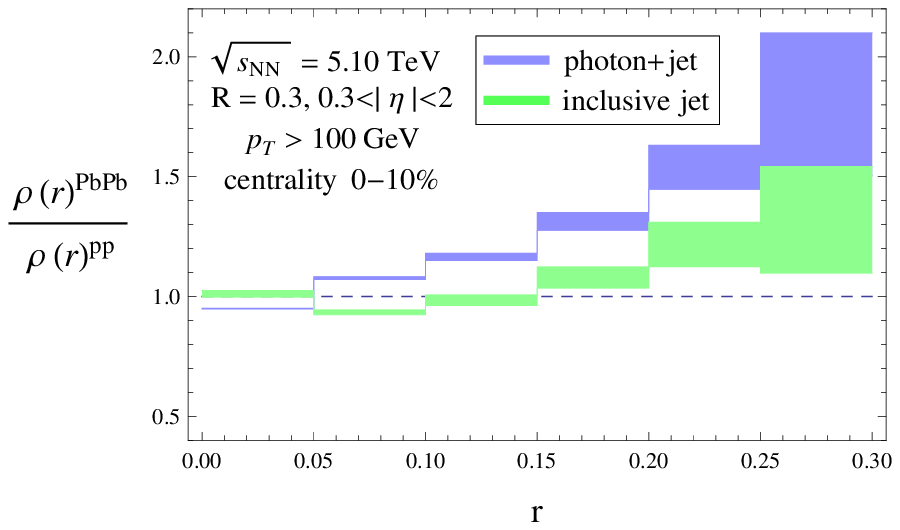}
\includegraphics[width=0.49\textwidth]{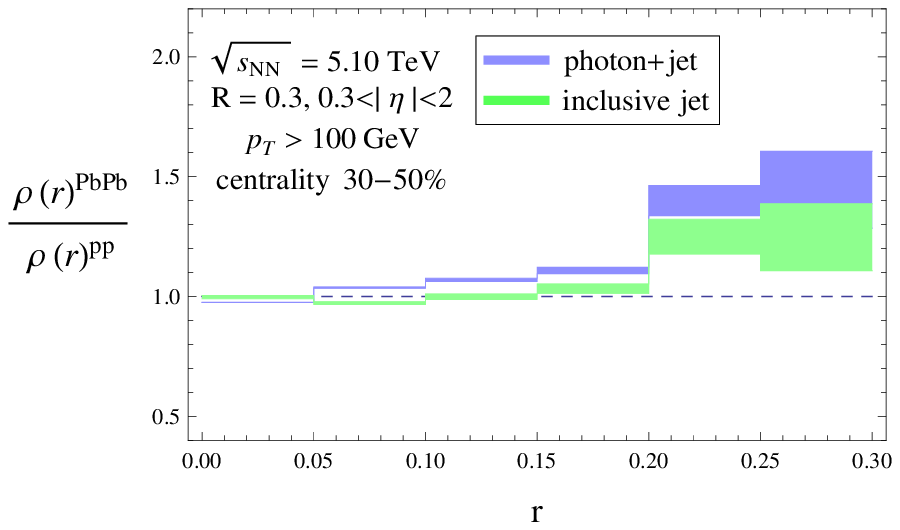}
\caption{Theoretical predictions of the modification of differential jet shapes for inclusive jets (green band) and photon-tagged jets (blue band), with $R=0.3$ in central (left panel) and mid-peripheral (right panel) Pb+Pb collisions at $\sqrt{s_{\rm NN}}\approx 5.1$~TeV at the LHC. We impose cuts on the jet transverse momentum $p_T > 100$~GeV and pseudo-rapidity $0.3<|\eta|<2.0$.  The coupling between the jet and the medium is fixed at $g=2.0$, and the bands correspond to the theoretical uncertainties estimated by varying the jet energy scales.}
\label{RdiffSHAPE_INCLvsPHOT_R0.3}
\end{figure}

\begin{figure}[!t!t]
\centering
\includegraphics[width=0.49\textwidth]{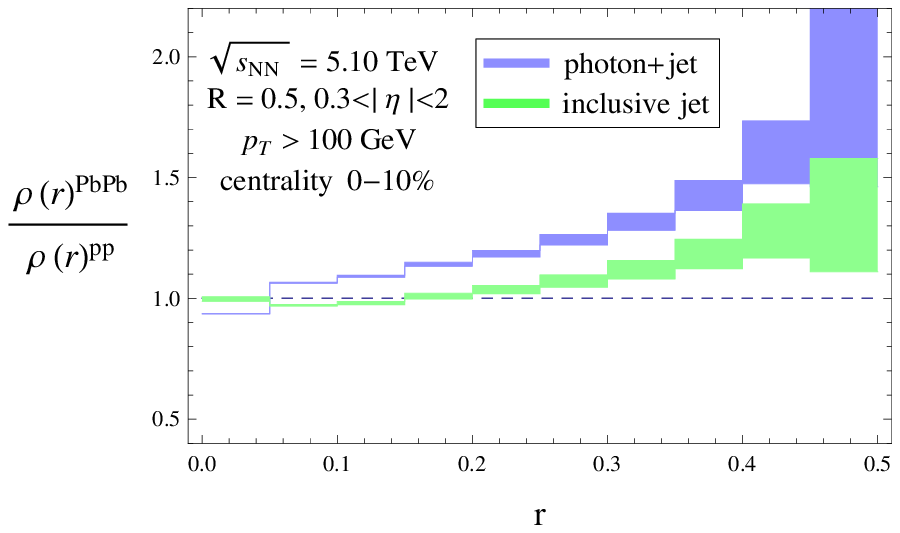}
\includegraphics[width=0.49\textwidth]{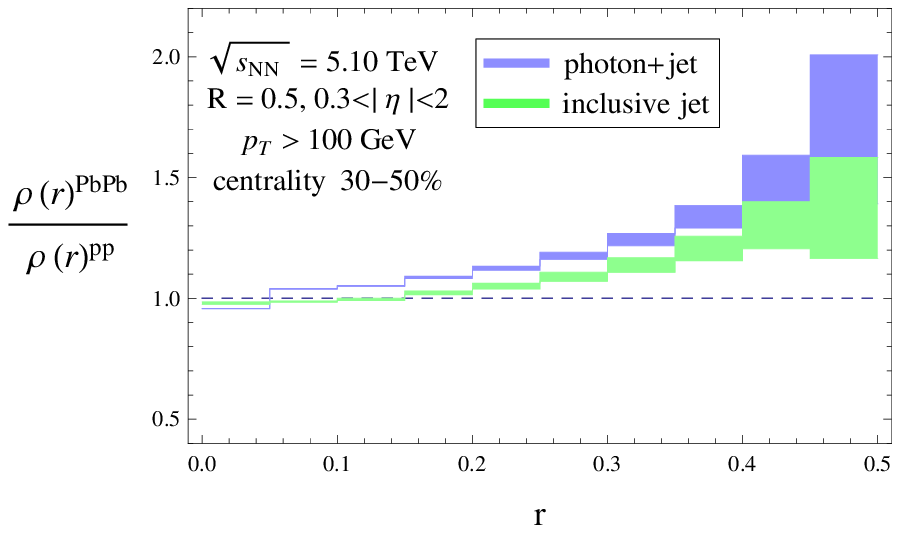}
\caption{Theoretical predictions of the modification of differential jet shapes for inclusive jets (green band) and photon-tagged jets (blue band), with $R=0.5$ in central (left panel) and mid-peripheral (right panel) Pb+Pb collisions at $\sqrt{s_{\rm NN}}\approx 5.1$~TeV at the LHC. The cuts of $p_T > 100$~GeV and $0.3<|\eta|<2.0$ are imposed, and the bands represent the theoretical uncertainties in the calculations.}
\label{RdiffSHAPE_INCLvsPHOT_R0.5}
\end{figure}

\section{Summary and discussion}
\label{sec:conc}

In this paper, using soft collinear effective theory  and its extension SCET$_{\rm G}$ to include collinear parton interactions
in dense QCD matter via Glauber gluon exchange, we calculate the modification of jet cross sections and differential jet shapes in lead-lead collisions at the LHC. The medium contributions to the jet energy function are evaluated using the medium-induced splitting kernels obtained in SCET$_{\rm G}$. This work emphasizes the consistent theoretical descriptions of hadron and jet observables in heavy ion collisions. The results presented in this paper also include cold nuclear matter (CNM) effects, which have been shown to affect the nuclear modification factor $R_{AA}$ and seen more clearly at high transverse momentum. We find that the calculations can describe well the centrality and the $p_T$ dependence of the inclusive jet suppression observed by the ALICE, ATLAS and CMS collaborations in $\sqrt{s_{\rm NN}}=2.76$~TeV Pb+Pb collisions at the LHC, 
%with the coupling between the jet and the medium $g=2.0\pm0.2$ and small CNM effects. 
with the coupling between the jet and the medium $g=2.0\pm0.2$ and CNM effects with $\mu_{\textcolor{\highlightA}{\rm CNM}}=0.35$ GeV. 
%The jet radius $R$ dependence of jet quenching is also consistent with the measurements.
The jet radius $R$ dependence of jet quenching is also qualitatively consistent with the measurements.

We also calculate the jet shape in nucleus-nucleus collisions. We find that the non-trivial behavior of the jet shape modification is caused by both the different quark and gluon jet cross section suppressions and the jet-by-jet broadening. The cross section of gluon-initiated jets is more suppressed, which enhances the fraction of quark-initiated jets having a narrower energy profile. This causes the attenuation of the jet shape in the mid $r$ region. On the other hand, the broadening of jets results in the enhancement of the jet shape near the periphery of the jet. The calculation provides for the first time a quantitative description of the jet shape modification in Pb+Pb collisions at the LHC. We would also like to emphasize that, the jet shape and substructure observables in general are \textcolor{\highlightA}{less sensitive} to initial state effects. This allows us to cleanly disentangle final state jet-medium interactions and directly probes the medium properties. A cleaner study of the initial state effects in A+A collisions can then be done.

For the upcoming LHC Run II measurements, we present theoretical predictions of the jet shape modification and the cross section suppression for inclusive jets and photon-tagged jets. We find that the cross section suppression at high $p_T$ can provide information about the cold nuclear matter effects. Since photon-tagged jets are predominately quark-initiated, the cross section is expected to be less suppressed compared to inclusive jets. On the other hand, the broadening of the photon-tagged jet is more apparent.

With the understanding of the transverse distribution of intra-jet particles through the studies of jet shapes and cross sections, it would be interesting to study the modification of their longitudinal distribution to gain orthogonal information about the in-medium parton shower~\cite{Borghini:2005em,Adare:2012qi,Chatrchyan:2014ava,Aad:2014wha}. We leave the studies of jet fragmentation function modification using the SCET formulation \cite{Procura:2009vm,Baumgart:2014upa,Ritzmann:2014mka}, as well as the modification of jet masses \cite{Dasgupta:2012hg,Chien:2012ur,Jouttenus:2013hs,Liu:2014oog} which probes the jet formation mechanism at the soft scale, for future work.

\section*{Acknowledgments}
Y.-T. C. would like to thank Doga Can Gulhan, Christopher Lee, Yen-Jie Lee, Emanuele Mereghetti, Daniel Pablos and Gregory Soyez for helpful discussions, and to Yen-Jie Lee for very useful comments on the manuscript. \textcolor{\highlight}{We would also like to thank the anonymous referee for many useful suggestions}. This work is supported by the US Department of Energy, Office of Science under Contract Nos. DE-AC52-06NA25396 and by the DOE Early Career Program.

\bibliographystyle{JHEP3}
\bibliography{jet_shape_1}

\providecommand{\href}[2]{#2}\begingroup\raggedright\begin{thebibliography}{100}

\bibitem{Gyulassy:1990ye}
M.~Gyulassy and M.~Plumer {\em Phys. Lett.} {\bf B243} (1990) 432--438.

\bibitem{Adcox:2001jp}
{\bf PHENIX} Collaboration, K.~Adcox {\em et~al.} {\em Phys. Rev. Lett.} {\bf
  88} (2002) 022301, [\href{http://arxiv.org/abs/nucl-ex/0109003}{{\tt
  nucl-ex/0109003}}].

\bibitem{Adler:2002xw}
{\bf STAR} Collaboration, C.~Adler {\em et~al.} {\em Phys. Rev. Lett.} {\bf 89}
  (2002) 202301, [\href{http://arxiv.org/abs/nucl-ex/0206011}{{\tt
  nucl-ex/0206011}}].

\bibitem{Adcox:2004mh}
{\bf PHENIX} Collaboration, K.~Adcox {\em et~al.} {\em Nucl. Phys.} {\bf A757}
  (2005) 184--283, [\href{http://arxiv.org/abs/nucl-ex/0410003}{{\tt
  nucl-ex/0410003}}].

\bibitem{Arsene:2004fa}
{\bf BRAHMS} Collaboration, I.~Arsene {\em et~al.} {\em Nucl. Phys.} {\bf A757}
  (2005) 1--27, [\href{http://arxiv.org/abs/nucl-ex/0410020}{{\tt
  nucl-ex/0410020}}].

\bibitem{Back:2004je}
B.~B. Back {\em et~al.} {\em Nucl. Phys.} {\bf A757} (2005) 28--101,
  [\href{http://arxiv.org/abs/nucl-ex/0410022}{{\tt nucl-ex/0410022}}].

\bibitem{Adams:2005dq}
{\bf STAR} Collaboration, J.~Adams {\em et~al.} {\em Nucl. Phys.} {\bf A757}
  (2005) 102--183, [\href{http://arxiv.org/abs/nucl-ex/0501009}{{\tt
  nucl-ex/0501009}}].

\bibitem{CMS:2012aa}
{\bf CMS} Collaboration, S.~Chatrchyan {\em et~al.} {\em Eur. Phys. J.} {\bf
  C72} (2012) 1945, [\href{http://arxiv.org/abs/1202.2554}{{\tt
  arXiv:1202.2554}}].

\bibitem{Aamodt:2010jd}
{\bf ALICE} Collaboration, K.~Aamodt {\em et~al.} {\em Phys. Lett.} {\bf B696}
  (2011) 30--39, [\href{http://arxiv.org/abs/1012.1004}{{\tt
  arXiv:1012.1004}}].

\bibitem{Abelev:2012hxa}
{\bf ALICE} Collaboration, B.~Abelev {\em et~al.} {\em Phys. Lett.} {\bf B720}
  (2013) 52--62, [\href{http://arxiv.org/abs/1208.2711}{{\tt
  arXiv:1208.2711}}].

\bibitem{Abelev:2013kqa}
{\bf ALICE} Collaboration, B.~Abelev {\em et~al.} {\em JHEP} {\bf 03} (2014)
  013, [\href{http://arxiv.org/abs/1311.0633}{{\tt arXiv:1311.0633}}].

\bibitem{Aad:2012vca}
{\bf ATLAS} Collaboration, G.~Aad {\em et~al.} {\em Phys.Lett.} {\bf B719}
  (2013) 220--241, [\href{http://arxiv.org/abs/1208.1967}{{\tt
  arXiv:1208.1967}}].

\bibitem{Aad:2014bxa}
{\bf ATLAS} Collaboration, G.~Aad {\em et~al.} {\em Phys. Rev. Lett.} {\bf 114}
  (2015), no.~7 072302, [\href{http://arxiv.org/abs/1411.2357}{{\tt
  arXiv:1411.2357}}].

\bibitem{Chatrchyan:2013kwa}
{\bf CMS} Collaboration, S.~Chatrchyan {\em et~al.} {\em Phys.Lett.} {\bf B730}
  (2014) 243--263, [\href{http://arxiv.org/abs/1310.0878}{{\tt
  arXiv:1310.0878}}].

\bibitem{CMS:prelim}
{\bf CMS} Collaboration, S.~Chatrchyan {\em et~al.}
  \href{http://arxiv.org/abs/CMS-HIN-12-004}{{\tt CMS-HIN-12-004}}.

\bibitem{Chatrchyan:2012gw}
{\bf CMS} Collaboration, S.~Chatrchyan {\em et~al.} {\em JHEP} {\bf 10} (2012)
  087, [\href{http://arxiv.org/abs/1205.5872}{{\tt arXiv:1205.5872}}].

\bibitem{Chatrchyan:2014ava}
{\bf CMS} Collaboration, S.~Chatrchyan {\em et~al.} {\em Phys. Rev.} {\bf C90}
  (2014), no.~2 024908, [\href{http://arxiv.org/abs/1406.0932}{{\tt
  arXiv:1406.0932}}].

\bibitem{Aad:2014wha}
{\bf ATLAS} Collaboration, G.~Aad {\em et~al.} {\em Phys. Lett.} {\bf B739}
  (2014) 320--342, [\href{http://arxiv.org/abs/1406.2979}{{\tt
  arXiv:1406.2979}}].

\bibitem{Chatrchyan:2013exa}
{\bf CMS} Collaboration, S.~Chatrchyan {\em et~al.} {\em Phys. Rev. Lett.} {\bf
  113} (2014), no.~13 132301, [\href{http://arxiv.org/abs/1312.4198}{{\tt
  arXiv:1312.4198}}]. [Erratum: Phys. Rev. Lett.115,no.2,029903(2015)].

\bibitem{Adam:2015ewa}
{\bf ALICE} Collaboration, J.~Adam {\em et~al.} {\em Phys. Lett.} {\bf B746}
  (2015) 1--14, [\href{http://arxiv.org/abs/1502.01689}{{\tt
  arXiv:1502.01689}}].

\bibitem{Chatrchyan:2012gt}
{\bf CMS} Collaboration, S.~Chatrchyan {\em et~al.} {\em Phys. Lett.} {\bf
  B718} (2013) 773--794, [\href{http://arxiv.org/abs/1205.0206}{{\tt
  arXiv:1205.0206}}].

\bibitem{Chatrchyan:2011sx}
{\bf CMS} Collaboration, S.~Chatrchyan {\em et~al.} {\em Phys.Rev.} {\bf C84}
  (2011) 024906, [\href{http://arxiv.org/abs/1102.1957}{{\tt
  arXiv:1102.1957}}].

\bibitem{Chatrchyan:2012nia}
{\bf CMS} Collaboration, S.~Chatrchyan {\em et~al.} {\em Phys. Lett.} {\bf
  B712} (2012) 176--197, [\href{http://arxiv.org/abs/1202.5022}{{\tt
  arXiv:1202.5022}}].

\bibitem{Aad:2010bu}
{\bf ATLAS} Collaboration, G.~Aad {\em et~al.} {\em Phys.Rev.Lett.} {\bf 105}
  (2010) 252303, [\href{http://arxiv.org/abs/1011.6182}{{\tt
  arXiv:1011.6182}}].

\bibitem{Aad:2013sla}
{\bf ATLAS} Collaboration, G.~Aad {\em et~al.} {\em Phys. Rev. Lett.} {\bf 111}
  (2013), no.~15 152301, [\href{http://arxiv.org/abs/1306.6469}{{\tt
  arXiv:1306.6469}}].

\bibitem{Adam:2015doa}
{\bf ALICE} Collaboration, J.~Adam {\em et~al.}
  \href{http://arxiv.org/abs/1506.03984}{{\tt arXiv:1506.03984}}.

\bibitem{Aad:2015bsa}
{\bf ATLAS} Collaboration, G.~Aad {\em et~al.}
  \href{http://arxiv.org/abs/1506.08656}{{\tt arXiv:1506.08656}}.

\bibitem{Gyulassy:1993hr}
M.~Gyulassy and X.-n. Wang {\em Nucl. Phys.} {\bf B420} (1994) 583--614,
  [\href{http://arxiv.org/abs/nucl-th/9306003}{{\tt nucl-th/9306003}}].

\bibitem{Wang:1994fx}
X.-N. Wang, M.~Gyulassy, and M.~Plumer {\em Phys. Rev.} {\bf D51} (1995)
  3436--3446, [\href{http://arxiv.org/abs/hep-ph/9408344}{{\tt
  hep-ph/9408344}}].

\bibitem{Zakharov:1996fv}
B.~G. Zakharov {\em JETP Lett.} {\bf 63} (1996) 952--957,
  [\href{http://arxiv.org/abs/hep-ph/9607440}{{\tt hep-ph/9607440}}].

\bibitem{Zakharov:1997uu}
B.~G. Zakharov {\em JETP Lett.} {\bf 65} (1997) 615--620,
  [\href{http://arxiv.org/abs/hep-ph/9704255}{{\tt hep-ph/9704255}}].

\bibitem{Baier:1996kr}
R.~Baier, Y.~L. Dokshitzer, A.~H. Mueller, S.~Peigne, and D.~Schiff {\em Nucl.
  Phys.} {\bf B483} (1997) 291--320,
  [\href{http://arxiv.org/abs/hep-ph/9607355}{{\tt hep-ph/9607355}}].

\bibitem{Baier:1998kq}
R.~Baier, Y.~L. Dokshitzer, A.~H. Mueller, and D.~Schiff {\em Nucl. Phys.} {\bf
  B531} (1998) 403--425, [\href{http://arxiv.org/abs/hep-ph/9804212}{{\tt
  hep-ph/9804212}}].

\bibitem{Gyulassy:2000er}
M.~Gyulassy, P.~Levai, and I.~Vitev {\em Nucl. Phys.} {\bf B594} (2001)
  371--419, [\href{http://arxiv.org/abs/nucl-th/0006010}{{\tt
  nucl-th/0006010}}].

\bibitem{Gyulassy:2000fs}
M.~Gyulassy, P.~Levai, and I.~Vitev {\em Phys.Rev.Lett.} {\bf 85} (2000)
  5535--5538, [\href{http://arxiv.org/abs/nucl-th/0005032}{{\tt
  nucl-th/0005032}}].

\bibitem{Wiedemann:2000za}
U.~A. Wiedemann {\em Nucl. Phys.} {\bf B588} (2000) 303--344,
  [\href{http://arxiv.org/abs/hep-ph/0005129}{{\tt hep-ph/0005129}}].

\bibitem{Wang:2001ifa}
X.-N. Wang and X.-f. Guo {\em Nucl. Phys.} {\bf A696} (2001) 788--832,
  [\href{http://arxiv.org/abs/hep-ph/0102230}{{\tt hep-ph/0102230}}].

\bibitem{Arnold:2001ba}
P.~B. Arnold, G.~D. Moore, and L.~G. Yaffe {\em JHEP} {\bf 11} (2001) 057,
  [\href{http://arxiv.org/abs/hep-ph/0109064}{{\tt hep-ph/0109064}}].

\bibitem{Arnold:2001ms}
P.~B. Arnold, G.~D. Moore, and L.~G. Yaffe {\em JHEP} {\bf 12} (2001) 009,
  [\href{http://arxiv.org/abs/hep-ph/0111107}{{\tt hep-ph/0111107}}].

\bibitem{Arnold:2002ja}
P.~B. Arnold, G.~D. Moore, and L.~G. Yaffe {\em JHEP} {\bf 06} (2002) 030,
  [\href{http://arxiv.org/abs/hep-ph/0204343}{{\tt hep-ph/0204343}}].

\bibitem{Casalderrey-Solana:2014bpa}
J.~Casalderrey-Solana, D.~C. Gulhan, J.~G. Milhano, D.~Pablos, and K.~Rajagopal
  {\em JHEP} {\bf 10} (2014) 19, [\href{http://arxiv.org/abs/1405.3864}{{\tt
  arXiv:1405.3864}}].

\bibitem{Gyulassy:2003mc}
M.~Gyulassy, I.~Vitev, X.-N. Wang, and B.-W. Zhang
  \href{http://arxiv.org/abs/nucl-th/0302077}{{\tt nucl-th/0302077}}.

\bibitem{Majumder:2010qh}
A.~Majumder and M.~Van~Leeuwen {\em Prog. Part. Nucl. Phys.} {\bf A66} (2011)
  41--92, [\href{http://arxiv.org/abs/1002.2206}{{\tt arXiv:1002.2206}}].

\bibitem{Mehtar-Tani:2013pia}
Y.~Mehtar-Tani, J.~G. Milhano, and K.~Tywoniuk {\em Int. J. Mod. Phys.} {\bf
  A28} (2013) 1340013, [\href{http://arxiv.org/abs/1302.2579}{{\tt
  arXiv:1302.2579}}].

\bibitem{CasalderreySolana:2011us}
J.~Casalderrey-Solana, H.~Liu, D.~Mateos, K.~Rajagopal, and U.~A. Wiedemann
  \href{http://arxiv.org/abs/1101.0618}{{\tt arXiv:1101.0618}}.

\bibitem{Bass:2008rv}
S.~A. Bass, C.~Gale, A.~Majumder, C.~Nonaka, G.-Y. Qin, T.~Renk, and J.~Ruppert
  {\em Phys. Rev.} {\bf C79} (2009) 024901,
  [\href{http://arxiv.org/abs/0808.0908}{{\tt arXiv:0808.0908}}].

\bibitem{Renk:2011aa}
T.~Renk {\em Phys. Rev.} {\bf C85} (2012) 044903,
  [\href{http://arxiv.org/abs/1112.2503}{{\tt arXiv:1112.2503}}].

\bibitem{Armesto:2011ht}
N.~Armesto {\em et~al.} {\em Phys. Rev.} {\bf C86} (2012) 064904,
  [\href{http://arxiv.org/abs/1106.1106}{{\tt arXiv:1106.1106}}].

\bibitem{Vitev:2008rz}
I.~Vitev, S.~Wicks, and B.-W. Zhang {\em JHEP} {\bf 0811} (2008) 093,
  [\href{http://arxiv.org/abs/0810.2807}{{\tt arXiv:0810.2807}}].

\bibitem{Gavai:2015pka}
R.~Gavai, A.~Jain, and R.~Sharma \href{http://arxiv.org/abs/1509.04671}{{\tt
  arXiv:1509.04671}}.

\bibitem{Sterman:1977wj}
G.~Sterman and S.~Weinberg {\em Phys. Rev. Lett.} {\bf 39} (1977) 1436.

\bibitem{Farhi:1977sg}
E.~Farhi {\em Phys.Rev.Lett.} {\bf 39} (1977) 1587--1588.

\bibitem{Georgi:1977sf}
H.~Georgi and M.~Machacek {\em Phys.Rev.Lett.} {\bf 39} (1977) 1237.

\bibitem{PhysRevLett.41.1581}
G.~C. Fox and S.~Wolfram {\em Phys. Rev. Lett.} {\bf 41} (Dec, 1978)
  1581--1585.

\bibitem{PhysRevLett.41.1585}
C.~L. Basham, L.~S. Brown, S.~D. Ellis, and S.~T. Love {\em Phys. Rev. Lett.}
  {\bf 41} (Dec, 1978) 1585--1588.

\bibitem{PhysRevD.19.2018}
C.~L. Basham, L.~S. Brown, S.~D. Ellis, and S.~T. Love {\em Phys. Rev. D} {\bf
  19} (Apr, 1979) 2018--2045.

\bibitem{Heister:2003aj}
{\bf ALEPH Collaboration} Collaboration, A.~Heister {\em et~al.} {\em
  Eur.Phys.J.} {\bf C35} (2004) 457--486.

\bibitem{Abdallah:2003xz}
{\bf DELPHI Collaboration} Collaboration, J.~Abdallah {\em et~al.} {\em
  Eur.Phys.J.} {\bf C29} (2003) 285--312,
  [\href{http://arxiv.org/abs/hep-ex/0307048}{{\tt hep-ex/0307048}}].

\bibitem{Achard:2004sv}
{\bf L3 Collaboration} Collaboration, P.~Achard {\em et~al.} {\em Phys.Rept.}
  {\bf 399} (2004) 71--174, [\href{http://arxiv.org/abs/hep-ex/0406049}{{\tt
  hep-ex/0406049}}].

\bibitem{Abbiendi:2004qz}
{\bf OPAL Collaboration} Collaboration, G.~Abbiendi {\em et~al.} {\em
  Eur.Phys.J.} {\bf C40} (2005) 287--316,
  [\href{http://arxiv.org/abs/hep-ex/0503051}{{\tt hep-ex/0503051}}].

\bibitem{Ellis:1990ek}
S.~D. Ellis, Z.~Kunszt, and D.~E. Soper {\em Phys. Rev. Lett.} {\bf 64} (1990)
  2121.

\bibitem{Ellis:1994dg}
S.~D. Ellis and D.~E. Soper {\em Phys. Rev. Lett.} {\bf 74} (1995) 5182--5185,
  [\href{http://arxiv.org/abs/hep-ph/9412342}{{\tt hep-ph/9412342}}].

\bibitem{Ellis:1992en}
S.~D. Ellis, Z.~Kunszt, and D.~E. Soper {\em Phys. Rev. Lett.} {\bf 69} (1992)
  1496--1499.

\bibitem{Becher:2008cf}
T.~Becher and M.~D. Schwartz {\em JHEP} {\bf 0807} (2008) 034,
  [\href{http://arxiv.org/abs/0803.0342}{{\tt arXiv:0803.0342}}].

\bibitem{Chien:2010kc}
Y.-T. Chien and M.~D. Schwartz {\em JHEP} {\bf 1008} (2010) 058,
  [\href{http://arxiv.org/abs/1005.1644}{{\tt arXiv:1005.1644}}].

\bibitem{Davison:2008vx}
R.~Davison and B.~Webber {\em Eur.Phys.J.} {\bf C59} (2009) 13--25,
  [\href{http://arxiv.org/abs/0809.3326}{{\tt arXiv:0809.3326}}].

\bibitem{Abbate:2010xh}
R.~Abbate, M.~Fickinger, A.~H. Hoang, V.~Mateu, and I.~W. Stewart {\em
  Phys.Rev.} {\bf D83} (2011) 074021,
  [\href{http://arxiv.org/abs/1006.3080}{{\tt arXiv:1006.3080}}].

\bibitem{Hoang:2015hka}
A.~H. Hoang, D.~W. Kolodrubetz, V.~Mateu, and I.~W. Stewart {\em Phys. Rev.}
  {\bf D91} (2015), no.~9 094018, [\href{http://arxiv.org/abs/1501.04111}{{\tt
  arXiv:1501.04111}}].

\bibitem{Gallicchio:2011xq}
J.~Gallicchio and M.~D. Schwartz {\em Phys.Rev.Lett.} {\bf 107} (2011) 172001,
  [\href{http://arxiv.org/abs/1106.3076}{{\tt arXiv:1106.3076}}].

\bibitem{Gallicchio:2012ez}
J.~Gallicchio and M.~D. Schwartz {\em JHEP} {\bf 1304} (2013) 090,
  [\href{http://arxiv.org/abs/1211.7038}{{\tt arXiv:1211.7038}}].

\bibitem{Dasgupta:2014yra}
M.~Dasgupta, F.~Dreyer, G.~P. Salam, and G.~Soyez {\em JHEP} {\bf 04} (2015)
  039, [\href{http://arxiv.org/abs/1411.5182}{{\tt arXiv:1411.5182}}].

\bibitem{Chien:2015cka}
Y.-T. Chien, A.~Hornig, and C.~Lee \href{http://arxiv.org/abs/1509.04287}{{\tt
  arXiv:1509.04287}}.

\bibitem{Ellis:1992qq}
S.~D. Ellis, Z.~Kunszt, and D.~E. Soper {\em Phys.Rev.Lett.} {\bf 69} (1992)
  3615--3618, [\href{http://arxiv.org/abs/hep-ph/9208249}{{\tt
  hep-ph/9208249}}].

\bibitem{Seymour:1997kj}
M.~Seymour {\em Nucl.Phys.} {\bf B513} (1998) 269--300,
  [\href{http://arxiv.org/abs/hep-ph/9707338}{{\tt hep-ph/9707338}}].

\bibitem{Li:2011hy}
H.-n. Li, Z.~Li, and C.-P. Yuan {\em Phys.Rev.Lett.} {\bf 107} (2011) 152001,
  [\href{http://arxiv.org/abs/1107.4535}{{\tt arXiv:1107.4535}}].

\bibitem{Li:2012bw}
H.-n. Li, Z.~Li, and C.-P. Yuan {\em Phys.Rev.} {\bf D87} (2013) 074025,
  [\href{http://arxiv.org/abs/1206.1344}{{\tt arXiv:1206.1344}}].

\bibitem{Vitev:2009rd}
I.~Vitev and B.-W. Zhang {\em Phys. Rev. Lett.} {\bf 104} (2010) 132001,
  [\href{http://arxiv.org/abs/0910.1090}{{\tt arXiv:0910.1090}}].

\bibitem{Salgado:2003rv}
C.~A. Salgado and U.~A. Wiedemann {\em Phys. Rev. Lett.} {\bf 93} (2004)
  042301, [\href{http://arxiv.org/abs/hep-ph/0310079}{{\tt hep-ph/0310079}}].

\bibitem{Renk:2009hv}
T.~Renk {\em Phys. Rev.} {\bf C80} (2009) 044904,
  [\href{http://arxiv.org/abs/0906.3397}{{\tt arXiv:0906.3397}}].

\bibitem{Ma:2013uqa}
G.-L. Ma {\em Phys.Rev.} {\bf C89} (2014) 024902,
  [\href{http://arxiv.org/abs/1309.5555}{{\tt arXiv:1309.5555}}].

\bibitem{Ramos:2014mba}
R.~Perez-Ramos and T.~Renk \href{http://arxiv.org/abs/1401.5283}{{\tt
  arXiv:1401.5283}}.

\bibitem{He:2011pd}
Y.~He, I.~Vitev, and B.-W. Zhang {\em Phys.Lett.} {\bf B713} (2012) 224--232,
  [\href{http://arxiv.org/abs/1105.2566}{{\tt arXiv:1105.2566}}].

\bibitem{Neufeld:2010fj}
R.~B. Neufeld, I.~Vitev, and B.~W. Zhang {\em Phys. Rev.} {\bf C83} (2011)
  034902, [\href{http://arxiv.org/abs/1006.2389}{{\tt arXiv:1006.2389}}].

\bibitem{Neufeld:2012df}
R.~B. Neufeld and I.~Vitev {\em Phys. Rev. Lett.} {\bf 108} (2012) 242001,
  [\href{http://arxiv.org/abs/1202.5556}{{\tt arXiv:1202.5556}}].

\bibitem{Dai:2012am}
W.~Dai, I.~Vitev, and B.-W. Zhang {\em Phys. Rev. Lett.} {\bf 110} (2013),
  no.~14 142001, [\href{http://arxiv.org/abs/1207.5177}{{\tt
  arXiv:1207.5177}}].

\bibitem{Stavreva:2012aa}
T.~Stavreva, F.~Arleo, and I.~Schienbein {\em JHEP} {\bf 02} (2013) 072,
  [\href{http://arxiv.org/abs/1211.6744}{{\tt arXiv:1211.6744}}].

\bibitem{Huang:2013vaa}
J.~Huang, Z.-B. Kang, and I.~Vitev {\em Phys. Lett.} {\bf B726} (2013)
  251--256, [\href{http://arxiv.org/abs/1306.0909}{{\tt arXiv:1306.0909}}].

\bibitem{Huang:2015mva}
J.~Huang, Z.-B. Kang, I.~Vitev, and H.~Xing
  \href{http://arxiv.org/abs/1505.03517}{{\tt arXiv:1505.03517}}.

\bibitem{Lokhtin:2011qq}
I.~P. Lokhtin, A.~V. Belyaev, and A.~M. Snigirev {\em Eur. Phys. J.} {\bf C71}
  (2011) 1650, [\href{http://arxiv.org/abs/1103.1853}{{\tt arXiv:1103.1853}}].

\bibitem{Young:2011qx}
C.~Young, B.~Schenke, S.~Jeon, and C.~Gale {\em Phys. Rev.} {\bf C84} (2011)
  024907, [\href{http://arxiv.org/abs/1103.5769}{{\tt arXiv:1103.5769}}].

\bibitem{Qin:2012gp}
G.-Y. Qin {\em Eur. Phys. J.} {\bf C74} (2014) 2959,
  [\href{http://arxiv.org/abs/1210.6610}{{\tt arXiv:1210.6610}}].

\bibitem{Wang:2013cia}
X.-N. Wang and Y.~Zhu {\em Phys. Rev. Lett.} {\bf 111} (2013), no.~6 062301,
  [\href{http://arxiv.org/abs/1302.5874}{{\tt arXiv:1302.5874}}].

\bibitem{Schenke:2009gb}
B.~Schenke, C.~Gale, and S.~Jeon {\em Phys. Rev.} {\bf C80} (2009) 054913,
  [\href{http://arxiv.org/abs/0909.2037}{{\tt arXiv:0909.2037}}].

\bibitem{Armesto:2009fj}
N.~Armesto, L.~Cunqueiro, and C.~A. Salgado {\em Eur. Phys. J.} {\bf C63}
  (2009) 679--690, [\href{http://arxiv.org/abs/0907.1014}{{\tt
  arXiv:0907.1014}}].

\bibitem{Zapp:2013vla}
K.~C. Zapp {\em Eur. Phys. J.} {\bf C74} (2014), no.~2 2762,
  [\href{http://arxiv.org/abs/1311.0048}{{\tt arXiv:1311.0048}}].

\bibitem{Zapp:2013zya}
K.~C. Zapp {\em Phys. Lett.} {\bf B735} (2014) 157--163,
  [\href{http://arxiv.org/abs/1312.5536}{{\tt arXiv:1312.5536}}].

\bibitem{Chien:2014nsa}
Y.-T. Chien and I.~Vitev {\em JHEP} {\bf 12} (2014) 061,
  [\href{http://arxiv.org/abs/1405.4293}{{\tt arXiv:1405.4293}}].

\bibitem{Chien:2014zna}
Y.-T. Chien {\em Int. J. Mod. Phys. Conf. Ser.} {\bf 37} (2015) 1560047,
  [\href{http://arxiv.org/abs/1411.0741}{{\tt arXiv:1411.0741}}].

\bibitem{Bauer:2000ew}
C.~W. Bauer, S.~Fleming, and M.~E. Luke {\em Phys.Rev.} {\bf D63} (2000)
  014006, [\href{http://arxiv.org/abs/hep-ph/0005275}{{\tt hep-ph/0005275}}].

\bibitem{Bauer:2000yr}
C.~W. Bauer, S.~Fleming, D.~Pirjol, and I.~W. Stewart {\em Phys.Rev.} {\bf D63}
  (2001) 114020, [\href{http://arxiv.org/abs/hep-ph/0011336}{{\tt
  hep-ph/0011336}}].

\bibitem{Bauer:2001ct}
C.~W. Bauer and I.~W. Stewart {\em Phys.Lett.} {\bf B516} (2001) 134--142,
  [\href{http://arxiv.org/abs/hep-ph/0107001}{{\tt hep-ph/0107001}}].

\bibitem{Bauer:2001yt}
C.~W. Bauer, D.~Pirjol, and I.~W. Stewart {\em Phys.Rev.} {\bf D65} (2002)
  054022, [\href{http://arxiv.org/abs/hep-ph/0109045}{{\tt hep-ph/0109045}}].

\bibitem{Bauer:2002nz}
C.~W. Bauer, S.~Fleming, D.~Pirjol, I.~Z. Rothstein, and I.~W. Stewart {\em
  Phys.Rev.} {\bf D66} (2002) 014017,
  [\href{http://arxiv.org/abs/hep-ph/0202088}{{\tt hep-ph/0202088}}].

\bibitem{Beneke:2002ph}
M.~Beneke, A.~P. Chapovsky, M.~Diehl, and T.~Feldmann {\em Nucl. Phys.} {\bf
  B643} (2002) 431--476, [\href{http://arxiv.org/abs/hep-ph/0206152}{{\tt
  hep-ph/0206152}}].

\bibitem{Idilbi:2008vm}
A.~Idilbi and A.~Majumder {\em Phys.Rev.} {\bf D80} (2009) 054022,
  [\href{http://arxiv.org/abs/0808.1087}{{\tt arXiv:0808.1087}}].

\bibitem{Ovanesyan:2011xy}
G.~Ovanesyan and I.~Vitev {\em JHEP} {\bf 1106} (2011) 080,
  [\href{http://arxiv.org/abs/1103.1074}{{\tt arXiv:1103.1074}}].

\bibitem{Ovanesyan:2011kn}
G.~Ovanesyan and I.~Vitev {\em Phys.Lett.} {\bf B706} (2012) 371--378,
  [\href{http://arxiv.org/abs/1109.5619}{{\tt arXiv:1109.5619}}].

\bibitem{Fickinger:2013xwa}
M.~Fickinger, G.~Ovanesyan, and I.~Vitev {\em JHEP} {\bf 1307} (2013) 059,
  [\href{http://arxiv.org/abs/1304.3497}{{\tt arXiv:1304.3497}}].

\bibitem{Kang:2014xsa}
Z.-B. Kang, R.~Lashof-Regas, G.~Ovanesyan, P.~Saad, and I.~Vitev
  \href{http://arxiv.org/abs/1405.2612}{{\tt arXiv:1405.2612}}.

\bibitem{Chien:2015vja}
Y.-T. Chien, A.~Emerman, Z.-B. Kang, G.~Ovanesyan, and I.~Vitev
  \href{http://arxiv.org/abs/1509.02936}{{\tt arXiv:1509.02936}}.

\bibitem{Vitev:2005yg}
I.~Vitev {\em Phys. Lett.} {\bf B630} (2005) 78--84,
  [\href{http://arxiv.org/abs/hep-ph/0501255}{{\tt hep-ph/0501255}}].

\bibitem{Freedman:2011kj}
S.~M. Freedman and M.~Luke {\em Phys.Rev.} {\bf D85} (2012) 014003,
  [\href{http://arxiv.org/abs/1107.5823}{{\tt arXiv:1107.5823}}].

\bibitem{Feige:2012vc}
I.~Feige, M.~D. Schwartz, I.~W. Stewart, and J.~Thaler {\em Phys.Rev.Lett.}
  {\bf 109} (2012) 092001, [\href{http://arxiv.org/abs/1204.3898}{{\tt
  arXiv:1204.3898}}].

\bibitem{Feige:2013zla}
I.~Feige and M.~D. Schwartz {\em Phys.Rev.} {\bf D88} (2013), no.~6 065021,
  [\href{http://arxiv.org/abs/1306.6341}{{\tt arXiv:1306.6341}}].

\bibitem{Becher:2006qw}
T.~Becher and M.~Neubert {\em Phys.Lett.} {\bf B637} (2006) 251--259,
  [\href{http://arxiv.org/abs/hep-ph/0603140}{{\tt hep-ph/0603140}}].

\bibitem{Becher:2010pd}
T.~Becher and G.~Bell {\em Phys.Lett.} {\bf B695} (2011) 252--258,
  [\href{http://arxiv.org/abs/1008.1936}{{\tt arXiv:1008.1936}}].

\bibitem{Gaunt:2014xxa}
J.~R. Gaunt and M.~Stahlhofen {\em JHEP} {\bf 1412} (2014) 146,
  [\href{http://arxiv.org/abs/1409.8281}{{\tt arXiv:1409.8281}}].

\bibitem{Gaunt:2014xga}
J.~R. Gaunt, M.~Stahlhofen, and F.~J. Tackmann {\em JHEP} {\bf 1404} (2014)
  113, [\href{http://arxiv.org/abs/1401.5478}{{\tt arXiv:1401.5478}}].

\bibitem{Gaunt:2014cfa}
J.~Gaunt, M.~Stahlhofen, and F.~J. Tackmann {\em JHEP} {\bf 1408} (2014) 020,
  [\href{http://arxiv.org/abs/1405.1044}{{\tt arXiv:1405.1044}}].

\bibitem{Ritzmann:2014mka}
M.~Ritzmann and W.~J. Waalewijn {\em Phys. Rev.} {\bf D90} (2014), no.~5
  054029, [\href{http://arxiv.org/abs/1407.3272}{{\tt arXiv:1407.3272}}].

\bibitem{Miller:2007ri}
M.~L. Miller, K.~Reygers, S.~J. Sanders, and P.~Steinberg {\em Ann. Rev. Nucl.
  Part. Sci.} {\bf 57} (2007) 205--243,
  [\href{http://arxiv.org/abs/nucl-ex/0701025}{{\tt nucl-ex/0701025}}].

\bibitem{Vitev:2007ve}
I.~Vitev {\em Phys. Rev.} {\bf C75} (2007) 064906,
  [\href{http://arxiv.org/abs/hep-ph/0703002}{{\tt hep-ph/0703002}}].

\bibitem{Kang:2015mta}
Z.-B. Kang, I.~Vitev, and H.~Xing \href{http://arxiv.org/abs/1507.05987}{{\tt
  arXiv:1507.05987}}.

\bibitem{Adare:2015gla}
A.~Adare {\em et~al.} \href{http://arxiv.org/abs/1509.04657}{{\tt
  arXiv:1509.04657}}.

\bibitem{ATLAS:2014cpa}
{\bf ATLAS} Collaboration, G.~Aad {\em et~al.} {\em Phys. Lett.} {\bf B748}
  (2015) 392--413, [\href{http://arxiv.org/abs/1412.4092}{{\tt
  arXiv:1412.4092}}].

\bibitem{Kang:2012kc}
Z.-B. Kang, I.~Vitev, and H.~Xing {\em Phys. Lett.} {\bf B718} (2012) 482--487,
  [\href{http://arxiv.org/abs/1209.6030}{{\tt arXiv:1209.6030}}].

\bibitem{Chatrchyan:2014hqa}
{\bf CMS} Collaboration, S.~Chatrchyan {\em et~al.} {\em Eur. Phys. J.} {\bf
  C74} (2014), no.~7 2951, [\href{http://arxiv.org/abs/1401.4433}{{\tt
  arXiv:1401.4433}}].

\bibitem{Cacciari:2008gp}
M.~Cacciari, G.~P. Salam, and G.~Soyez {\em JHEP} {\bf 0804} (2008) 063,
  [\href{http://arxiv.org/abs/0802.1189}{{\tt arXiv:0802.1189}}].

\bibitem{Tung:2006tb}
W.~Tung, H.~Lai, A.~Belyaev, J.~Pumplin, D.~Stump, {\em et~al.} {\em JHEP} {\bf
  0702} (2007) 053, [\href{http://arxiv.org/abs/hep-ph/0611254}{{\tt
  hep-ph/0611254}}].

\bibitem{Neufeld:2011yh}
R.~B. Neufeld and I.~Vitev {\em Phys. Rev.} {\bf C86} (2012) 024905,
  [\href{http://arxiv.org/abs/1105.2067}{{\tt arXiv:1105.2067}}].

\bibitem{Borghini:2005em}
N.~Borghini and U.~A. Wiedemann \href{http://arxiv.org/abs/hep-ph/0506218}{{\tt
  hep-ph/0506218}}.

\bibitem{Adare:2012qi}
{\bf PHENIX} Collaboration, A.~Adare {\em et~al.} {\em Phys. Rev. Lett.} {\bf
  111} (2013), no.~3 032301, [\href{http://arxiv.org/abs/1212.3323}{{\tt
  arXiv:1212.3323}}].

\bibitem{Procura:2009vm}
M.~Procura and I.~W. Stewart {\em Phys. Rev.} {\bf D81} (2010) 074009,
  [\href{http://arxiv.org/abs/0911.4980}{{\tt arXiv:0911.4980}}]. [Erratum:
  Phys. Rev.D83,039902(2011)].

\bibitem{Baumgart:2014upa}
M.~Baumgart, A.~K. Leibovich, T.~Mehen, and I.~Z. Rothstein {\em JHEP} {\bf 11}
  (2014) 003, [\href{http://arxiv.org/abs/1406.2295}{{\tt arXiv:1406.2295}}].

\bibitem{Dasgupta:2012hg}
M.~Dasgupta, K.~Khelifa-Kerfa, S.~Marzani, and M.~Spannowsky {\em JHEP} {\bf
  10} (2012) 126, [\href{http://arxiv.org/abs/1207.1640}{{\tt
  arXiv:1207.1640}}].

\bibitem{Chien:2012ur}
Y.-T. Chien, R.~Kelley, M.~D. Schwartz, and H.~X. Zhu {\em Phys. Rev.} {\bf
  D87} (2013), no.~1 014010, [\href{http://arxiv.org/abs/1208.0010}{{\tt
  arXiv:1208.0010}}].

\bibitem{Jouttenus:2013hs}
T.~T. Jouttenus, I.~W. Stewart, F.~J. Tackmann, and W.~J. Waalewijn {\em Phys.
  Rev.} {\bf D88} (2013), no.~5 054031,
  [\href{http://arxiv.org/abs/1302.0846}{{\tt arXiv:1302.0846}}].

\bibitem{Liu:2014oog}
Z.~L. Liu, C.~S. Li, J.~Wang, and Y.~Wang {\em JHEP} {\bf 04} (2015) 005,
  [\href{http://arxiv.org/abs/1412.1337}{{\tt arXiv:1412.1337}}].

\end{thebibliography}\endgroup

\end{document}